\documentclass[aps,pre,onecolumn,superscriptaddress]{revtex4-1}
\usepackage{graphicx}
\usepackage{amssymb}
\usepackage{amsmath}
\usepackage{times}
\usepackage{tabularx}
\usepackage{color}
\usepackage{subfigure}
\usepackage{natbib}
\DeclareMathOperator{\erf}{erf}
\DeclareMathOperator{\erfc}{erfc}

\graphicspath{{./Figures/}}

\makeindex
\begin{document}
\newcommand{\lbra}{{_{L}\langle}} \newcommand{\lket}{{\rangle}_{L}}
\newcommand{\mbra}{{_{M}\langle}} \newcommand{\mket}{{\rangle}_{M}}
\newcommand{\bra}{{\langle}} \newcommand{\ket}{{\rangle}}
\newcommand{\mrm}[1]{\mathrm{#1}}
\newcommand{\mbf}[1]{\mbox{\boldmath$\mathrm{#1}$}}
\newcommand{\bfsub}[1]{\mbox{\boldmath${\scriptstyle\mathrm{#1}}$}}
\newcommand{\rmsub}[1]{\mbox{${\scriptstyle\mathrm{#1}}$}}
\renewcommand{\baselinestretch}{1.2}
\newcommand{\skipline}{\par\vspace{8pt}}
\newcommand{\skiphalfline}{\par\vspace{4pt}}
\newcommand{\HRule}{\rule{\linewidth}{0.2mm}}

\renewcommand{\thefootnote}{\fnsymbol{footnote}}

\newcommand{\tens}[1]{\underline{\underline{#1}}}
\newcommand{\vect}[1]{\mathbf{#1}}
\newcommand{\avect}[1]{\underline{#1}}
\newcommand{\laplacian}{\vec{\nabla}^2}
\newcommand{\del}{\nabla}
\newcommand{\diverg}{\vec{\nabla}\cdot}
\newcommand{\nndist}{\left<L_{nn}\right>}
\newcommand{\singlet}[1]{\rho^{(1)}\left(\vect{#1}\right)}
\newcommand{\doublet}[2]{\rho^{(2)}\left(\vect{#1},\vect{#2}\right)}
\newcommand{\condsinglet}[2]{\rho\left(\vect{#1} \middle\vert \vect{#2}\right)}
\newcommand{\rhocond}[2]{\rho\left(\vect{#1} \middle\vert \vect{#2}\right)}
\newcommand{\vdiff}[2]{\left|\vect{#1} - \vect{#2}\right|}
\newcommand{\ie}{\emph{i.e.}}
\newcommand{\eg}{\emph{e.g.}}
\newcommand{\etc}{\emph{etc.}}
\newcommand{\rhointracond}{\varrho_{M|\alpha}(\vect{R_M}|\vect{r})}

\newcommand{\phiR}{\ensuremath{\phi_R}}
\newcommand{\phis}{\ensuremath{\phi_0}}
\newcommand{\phil}{\ensuremath{\phi_1}}
\newcommand{\phiRl}{\ensuremath{\phi_{R1}}}
\newcommand{\V}{\ensuremath{\mathcal{V}}}
\newcommand{\Vr}{\ensuremath{\mathcal{V}_R}}
\newcommand{\Vs}{\ensuremath{\mathcal{V}_0}}
\newcommand{\Vl}{\ensuremath{\mathcal{V}_1}}
\newcommand{\Vrl}{\ensuremath{\mathcal{V}_{R1}}}
\newcommand{\uLJ}{\ensuremath{u_{\text{LJ}}}}
\newcommand{\sLJ}{\ensuremath{\sigma_{\text{LJ}}}}
\newcommand{\rnot}{\ensuremath{r_0}}
\newcommand{\eLJ}{\ensuremath{\varepsilon_{\text{LJ}}}}
\newcommand{\lmf}{\text{LMF}}
\newcommand{\sca}{\text{SCA}}
\newcommand{\pb}{\text{PB}}
\newcommand{\mpb}{\text{MPB}}
\newcommand{\lj}{\text{LJ}}
\newcommand{\sig}{\ensuremath{\sigma}}
\newcommand{\sigmin}{\ensuremath{\sigma_{\text{min}}}}
\newcommand{\vs}{\ensuremath{v_0(r)}}
\newcommand{\vl}{\ensuremath{v_1(r)}}
\newcommand{\us}{\ensuremath{u_0(r)}}
\newcommand{\ul}{\ensuremath{u_1(r)}}
\newcommand{\wca}{\text{WCA}}
\newcommand{\uwca}{\ensuremath{u_{\text{WCA}}(r)}}
\newcommand{\ua}{\ensuremath{u_{\text{attr}}(r)}}
\newcommand{\lB}{\ensuremath{L_B}}
\newcommand{\lG}{\ensuremath{L_G}}
\newcommand{\lw}{\ensuremath{L_w}}
\newcommand{\spce}{\text{SPC/E}}
\newcommand{\md}{\text{MD}}
\newcommand{\mc}{\text{MC}}
\newcommand{\nvt}{\text{NVT}}
\newcommand{\conddens}{\ensuremath{\condsinglet{\vect{r}^\prime}{\vect{r}}}}
\newcommand{\singdens}{\ensuremath{\singlet{\vect{r}}}}
\newcommand{\twodens}{\ensuremath{\doublet{\vect{r}}{\vect{r}^\prime}}}
\newcommand{\moldens}{\ensuremath{\rho_{\text{M}}(\vect{R}_{\text{M}})}}
\newcommand{\rhoqs}{\ensuremath{\rho^{q_\sigma}}}
\newcommand{\rhoq}{\ensuremath{\rho^q}}
\newcommand{\rhoG}{\ensuremath{\rho_G}}
\newcommand{\conv}{\ensuremath{\ast}}
\newcommand{\ybg}{\text{YBG}}
\newcommand{\hamiltonian}{\ensuremath{\mathcal{U}}}
\newcommand{\lagrangian}{\ensuremath{\mathcal{L}}}
\newcommand{\lcw}{\text{LCW}}
\newcommand{\hlr}{\text{HLR}}
\newcommand{\ha}{\text{HA}}
\newcommand{\ips}{\text{IPS}}
\newcommand{\Posm}{\ensuremath{P_{\text{osm}}}}
\newcommand{\huck}{\text{DH}}
\newcommand{\epm}{\text{EPM}}
\newcommand{\gsame}{\ensuremath{g_{\text{same}}}}
\newcommand{\gopp}{\ensuremath{g_{\text{opp}}}}
\newcommand{\Etot}{\ensuremath{E_{\text{tot}}}}
\newcommand{\Epol}{\ensuremath{E_{\text{pol}}}}
\newcommand{\Phipol}{\ensuremath{\Phi_{\text{pol}}}}
\newcommand{\tot}{\ensuremath{\text{tot}}}


\title {On the efficient and accurate short-ranged simulations of uniform polar molecular liquids}
\renewcommand{\thefootnote}{\fnsymbol{footnote}}
\author{Jocelyn M. Rodgers$^{2,}$\footnote{Corresponding author. Email: jrodgers@berkeley.edu}, Zhonghan Hu$^{1,}$\footnote{Corresponding author. Email: zhonghanhu@jlu.edu.cn}, and John D. Weeks$^{3,}$\footnote{Corresponding author. Email: jdw@umd.edu}\\
\vspace{6pt} 
$^{1}$State Key Laboratory of Supramolecular  Structure and Materials and Institute of Theoretical Chemistry, Jilin University, Changchun, 130012, China \\
$^{2}$ Physical Biosciences Division, Lawrence Berkeley National Laboratory, Berkeley, CA 94720 \\
$^{3}$ Institute for Physical Science and Technology and Department of Chemistry and Biochemistry, 
University of Maryland,  College Park, Maryland 20742}

\date{\today}

\begin{abstract}
We show that spherical truncations of the $1/r$ interactions in models for water and acetonitrile yield very accurate results in bulk simulations for all site-site pair correlation functions
as well as dipole-dipole correlation functions.
This good performance in bulk simulations
contrasts with the generally poor results found with the use of 
such truncations in nonuniform molecular systems.
We argue that Local Molecular Field (LMF) theory provides a
general theoretical framework that
gives the necessary corrections to simple truncations
in most nonuniform environments and explains the accuracy of spherical
truncations in uniform environments by showing that these corrections are very small.
LMF theory is derived from the exact Yvon-Born-Green (YBG) hierarchy by
making physically-motivated and well-founded approximations.
New and technically interesting derivations of both the YBG hierarchy
and LMF theory for a variety of site-site
molecular models are presented in appendices. The
main paper focuses on understanding the accuracy of these spherical
truncations in uniform systems both phenomenologically and quantitatively using LMF
theory.
\end{abstract}
\maketitle

\section{Introduction \label{sxn:intro}}

Spherical truncations of the Coulomb interactions present in
typical molecular models such as
CHARMM~\cite{MacKerellBashfordBellott.1998.All-Atom-Empirical-Potential-for-Molecular-Modeling,BrooksBrooksMackerell.2009.CHARMM:-The-Biomolecular-Simulation-Program}
and AMBER~\cite{DuanWuChowdhury.2003.A-point-charge-force-field-for-molecular-mechanics}
have long been used to keep computational cost in
check.  This cost in the simulation of
large biomolecules is compounded by the use of explicit water models
containing point charges to describe the hydrogen-bonding network and dielectric behavior of the
solvating water molecules. Since traditional particle-mesh Ewald sum treatments of
Coulomb interactions do not scale well in massively parallel
simulations, a computationally
compelling case can be made for the use of spherical truncations~\cite{Schulz:2009p5308}.
However, spherical truncations have been
shown to be clearly wrong when applied \emph{naively} in a variety of
nonuniform environments~\cite{Feller:1996p5303,Spohr.1997.Effect-of-Electrostatic-Boundary-Conditions-and-System}. 
For this reason, the use of short-ranged truncations of
$1/r$ interactions is typically viewed as an unjustified approximation.

There have been many attempts to place the use of
spherical truncations of $1/r$ on a more solid theoretical footing,
including site-site reaction field methods~\cite{HummerSoumpasisNeumann.1994.Computer-simulation-of-aqueous-Na-Cl-Electrolytes},
Wolf
summation~\cite{FennellGezelter.2006.Is-the-Ewald-summation-still-necessary-Pairwise,WolfKeblinskiPhillpot.1999.Exact-method-for-the-simulation-of-Coulombic-systems},
and isotropic periodic
summation~\cite{WuBrooks.2005.Isotropic-periodic-sum:-A-method-for-the-calculation,WuBrooks.2008.Using-the-isotropic-periodic-sum-method-to-calculate,WuBrooks.2009.Isotropic-periodic-sum-of-electrostatic-interactions-for-polar}.
Despite this work, the virtues and defects of spherical truncations of $1/r$ in various
applications remains a subject of ongoing debate in the current
literature~\cite{Schulz:2009p5308,Patra:2003p5327,Patra:2004p5329,Patra:2006p5332}.

Our approach, local molecular field (LMF)
theory~\cite{ChenKaurWeeks.2004.Connecting-systems-with-short-and-long,RodgersWeeks.2008.Local-molecular-field-theory-for-the-treatment},
uses an effective single particle potential to
account for the averaged effects of the long-ranged interactions neglected in
typical spherical truncations. It
gives a theoretical basis for the use of simple truncations in some cases, and also provides a
physically suggestive path for correction when such truncations fail. Moreover, recent work
has established a very efficient and accurate numerical method
to determine the effective field in LMF theory
using a simple linear response
approach~\cite{HuWeeks.2010.Efficient-Solutions-of-Self-Consistent-Mean-Field}.

LMF theory for general nonuniform fluids is derived from the exact statistical mechanical Yvon-Born-Green
(YBG) hierarchy~\cite{HansenMcDonald.2006.Theory-of-Simple-Liquids,McQuarrie.2000.Statistical-Mechanics}
by making two
physically-motivated and well-founded approximations. These rely on the
ability of well-chosen truncated potentials to yield accurate nearest-neighbor
correlations and on the corresponding slowly-varying nature of the remaining
long-ranged parts of the full
potential~\cite{RodgersWeeks.2008.Local-molecular-field-theory-for-the-treatment}.
Previous work has shown that LMF theory corrects two well known
failures of spherical truncations of $1/r$ interactions:
\begin{itemize}
\item simulations using LMF theory yield correct charge density
  profiles for water confined between two
  walls~\cite{RodgersWeeks.2008.Interplay-of-local-hydrogen-bonding-and-long-ranged-dipolar}
  and for ions confined between charged
  plates~\cite{RodgersKaurChen.2006.Attraction-between-like-charged-walls:-Short-ranged},
  and
\item simple analytical corrections derived via LMF theory result in
  accurate energies and pressures for uniform ionic and molecular
systems~\cite{ChenKaurWeeks.2004.Connecting-systems-with-short-and-long,RodgersWeeks.2009.Accurate-thermodynamics-for-short-ranged-truncations-of-Coulomb}
treated with spherical truncations.
\end{itemize}

In this paper we employ LMF theory to illustrate and explain why spherical truncations of
$1/r$ can often be applied very successfully for determining the structure
and thermodynamics of \emph{uniform} molecular systems.  When LMF theory is applied to
charge-charge interactions, all $1/r$ interactions are split into short and long ranged
parts \vs\ and \vl, such that
\begin{equation}
  \frac{1}{r} = \vs + \vl \equiv \frac{\erfc(r/\sigma)}{r} + \frac{\erf(r/\sigma)}{r}.
  \label{eq:CoulombSplit}
\end{equation}
Here \vl\ is the electrostatic potential from a unit
Gaussian charge distribution with width $\sigma$,
and \vs\ corresponds to the potential from a point charge surrounded by
a neutralizing Gaussian charge
distribution~\cite{ChenKaurWeeks.2004.Connecting-systems-with-short-and-long}.
Thus \vs\ vanishes at distances $r$ much greater
than the ``smoothing length'' $\sigma$ and at small distances the force from \vs\ 
approaches that of the bare point charge, so \vs\ can be though of as a ``Coulomb core potential''.

In the simple strong coupling approximation (SCA) to the full LMF theory,
we assume that all effects from  the long-ranged interactions due
to \vl\ may be neglected.  Thus the SCA is in essence a spherical
truncation where all $1/r$ interactions are replaced by
the short-ranged \vs, with $\sigma$ setting the scale for the truncation
distance. In Section~\ref{sxn:SCAresults}, we emphasize
the accuracy of the SCA for uniform molecular systems,
presenting results for SPC/E water and
acetonitrile (CH$_3$CN), including the effect of varying the range of the
short-ranged truncation of $1/r$ as represented by $\sigma$ in equation (\ref{eq:CoulombSplit}).
These results can be appreciated independent of the underlying
LMF theory discussed in the rest of this paper.

Furthermore, we demonstrate that spherical truncations can lead to
highly accurate dipole-dipole correlations in uniform molecular
systems.  This surprising result is in sharp contrast to findings by
Nezbeda, who used a different molecular-based truncation
scheme~\cite{Nezbeda.2005.Towards-a-Unified-View-of-Fluids}, and we
shall explain our success later using the full LMF theory. Then in
Sections~\ref{sxn:LMFdiscussion} and~\ref{sxn:SCAsuccess}, we
formulate LMF theory for bulk uniform site-site molecular fluids and
discuss the success of these spherical truncations and the neglect of
long-ranged interactions using the LMF theory framework for the
simpler bulk water system.  The form of the derived LMF equation and
the necessary approximations make clear why spherical truncations
can often give accurate structure in uniform systems,
despite their invalidity in nonuniform systems.

Detailed derivations of LMF equations for various molecular models are
discussed in complementary appendices. Here we build on previous
derivations of LMF theory for simple atomic fluids,  and focus in particular on the
derivation of the LMF
equation for a uniform system of site-site molecules, described by the
Hamiltonian
\begin{equation}
 {\cal U}=
\sum_{i=1}^N 
\omega_{M}(\mathbf R_{i}) +\frac{1}{2}\sum_{i=1}^{N} \sum_{j=1}^N (1-\delta_{ij}) \sum_{\alpha=1}^n
\sum_{\xi=1}^n u_{\alpha\xi}(|{\mathbf r}_i^{(\alpha)}-{\mathbf r}^{(\xi)}_{j}|) .
\label{eq:uniformU}
\end{equation}
Here $\vect{R}_i$ describes the positions of all
sites within a molecule $i$ connected by a generalized bonding
potential $\omega_M(\vect{R}_i)$, and
$u_{\alpha\xi}(|{\mathbf r}_i^{(\alpha)}-{\mathbf r}^{(\xi)}_{j}|)$
describes the general pair interaction between two sites $\alpha$
and $\xi$ on two different molecules $i$ and $j$ as insured by the
$\delta_{ij}$ term in equation (\ref{eq:uniformU}).  In
Appendix~\ref{app:SimpleYBGandLMF}, we present a derivation of the LMF
equation used in previous work
for small site-site molecules in a general external field.  Then in Appendices~\ref{app:YBG}
and~\ref{app:LMFbulk}, we present the notationally more complex
derivations of both the exact YBG hierarchy and the LMF equation for a uniform
fluid composed of these site-site molecules.  Finally, in
Appendix~\ref{app:charmm}, we present an abbreviated derivation for
larger molecules described by CHARMM- or AMBER-like Hamiltonians,
thus supporting the validity of our conclusions for systems composed of
much larger molecules.

\section{Strong Coupling Approximation (SCA) Simulations of Water and
Acetonitrile \label{sxn:SCAresults}}

\begin{figure}
  \begin{center}
    \subfigure[SPC/E Water]{
      \includegraphics[height=1in]{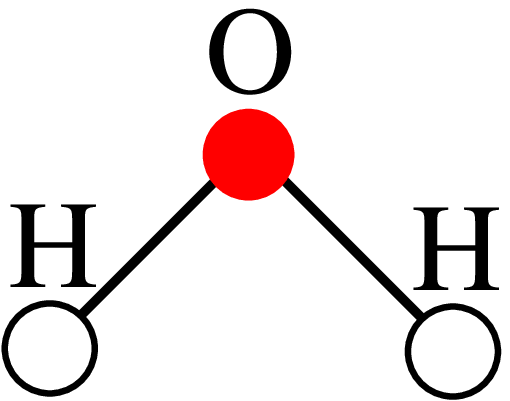}
    }
    \hspace{0.5in}
    \subfigure[Acetonitrile]{
      \includegraphics[height=1.2in]{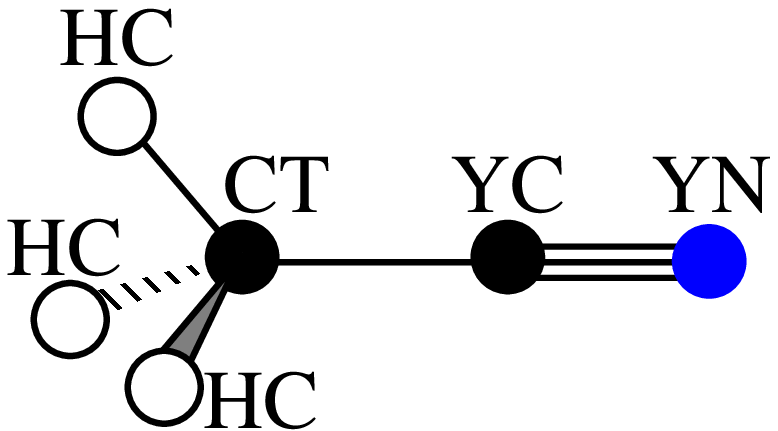}
    }
    \caption{Schematics depicting the geometry and site labels of the water
      model~\cite{BerendsenGrigeraStraatsma.1987.The-missing-term-in-effective-pair-potentials}
      and acetonitrile
      model~\cite{NikitinLyubartsev.2007.New-six-site-acetonitrile-model-for-simulations-of-liquid}
      used in this paper.}
    \label{fig:Models}
  \end{center}
\end{figure}

We present structural results for the simulation of two different
small site-site molecular models shown in Fig.~\ref{fig:Models}:
\begin{itemize}
\item SPC/E
  water~\cite{BerendsenGrigeraStraatsma.1987.The-missing-term-in-effective-pair-potentials},
  a rigid molecular model of a hydrogen-bonding fluid, and
\item acetonitrile, an AMBER-like flexible molecular
  model~\cite{NikitinLyubartsev.2007.New-six-site-acetonitrile-model-for-simulations-of-liquid}
  of a strongly dipolar fluid.
\end{itemize}
These models, along with annotation used for each site, are shown in
Fig.~\ref{fig:Models}.

For the water simulations, we present results for simulations of 1728
SPC/E water molecules in a cubic box of side length 37.27 \AA\ using
the {\sc dlpoly}2.16 simulation
package~\cite{SmithYongRodger.2002.DLPOLY:-Application-to-molecular-simulation}.  The system of water
molecules was equilibrated for 500 ps at 300 K using a Berendsen
thermostat with a time constant of 0.5 ps and a timestep of 1 fs.
Data was collected over the subsequent 1.5 ns.  Cutoff radii for the
potential ranged from 9.5 \AA\ up to 13.5 \AA\ for varying choices of
$\sigma$ in \vs.  The SCA simulations are compared to simulations using Ewald
summation with $\alpha=0.3$~\AA$^{-1}$ and $\vect{k}_{\rm
  max}=(10,10,10)$.

\begin{figure}
  \begin{center}
    \subfigure[$g_{\rm OO}(r)$]{
      \includegraphics[width=3.0in]{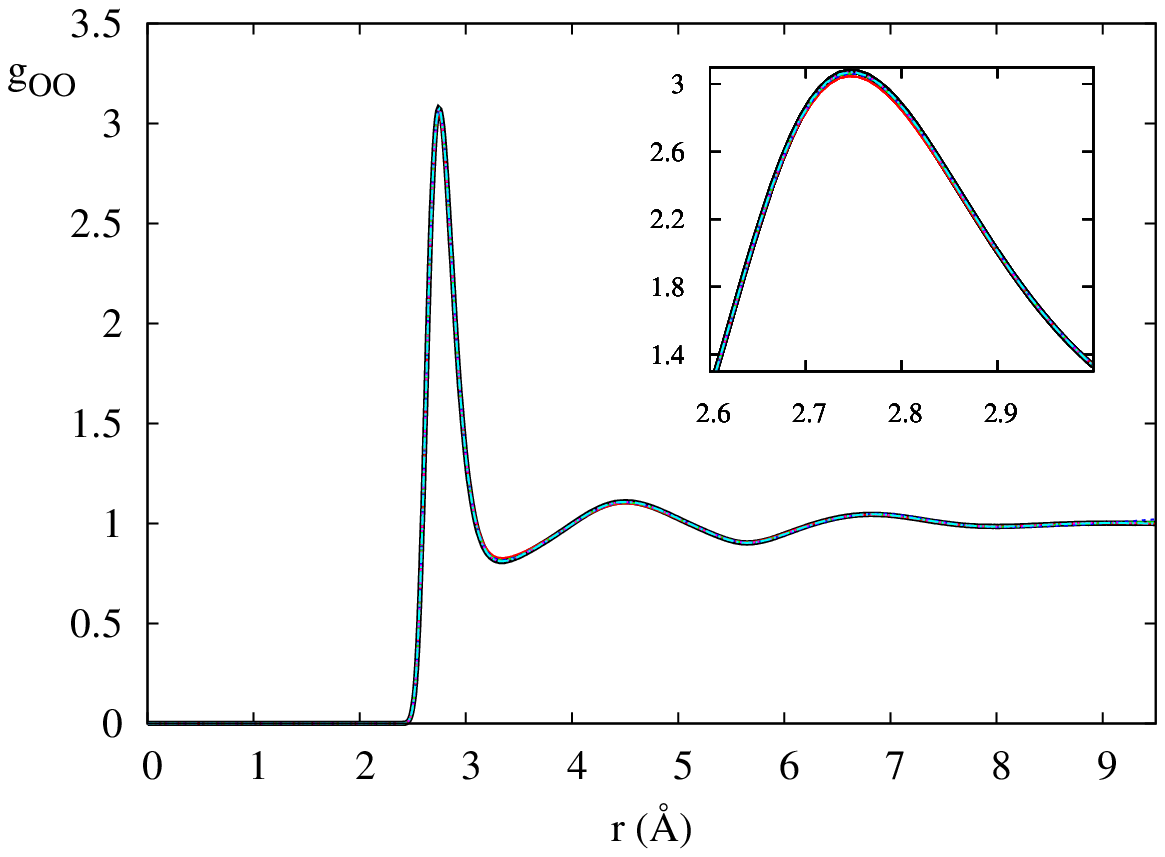}
    }
    \subfigure[$g_{\rm OH}(r)$]{
      \includegraphics[width=3.0in]{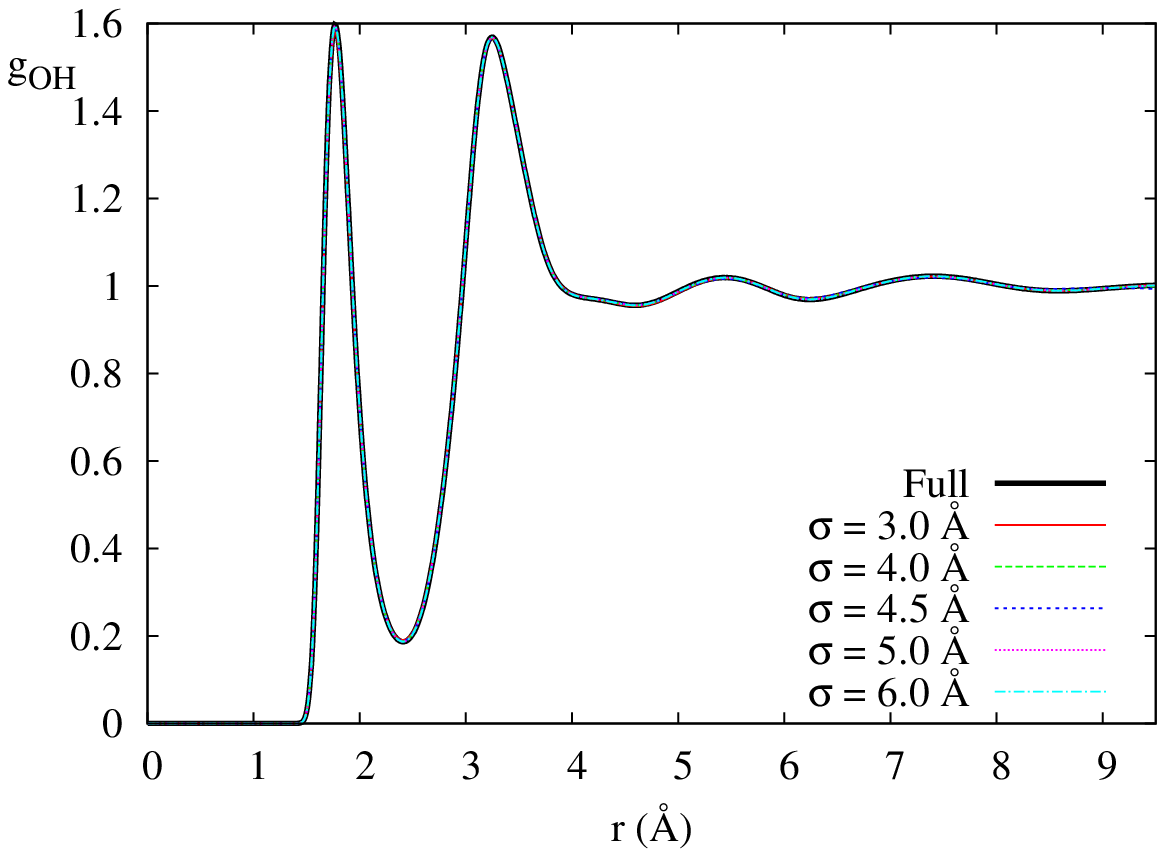}
    }
  \subfigure[$g_{\rm HH}(r)$]{
      \includegraphics[width=3.0in]{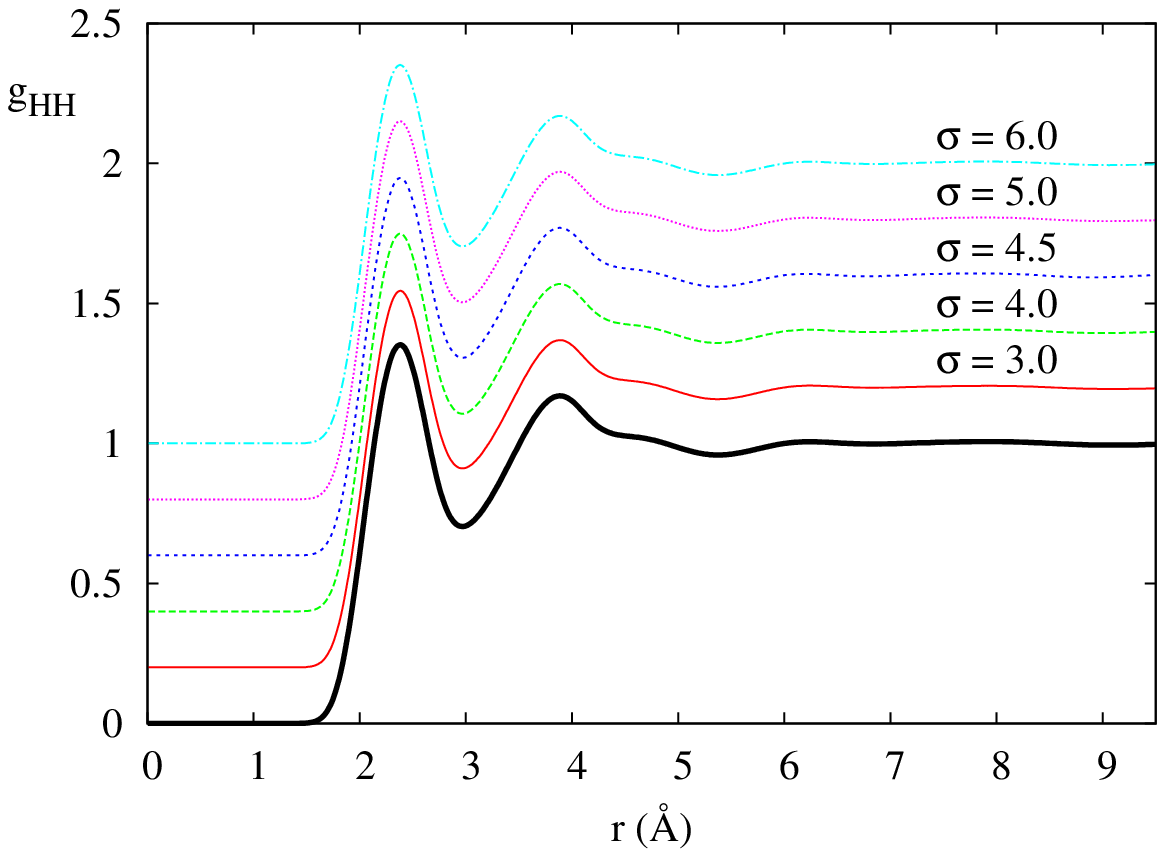}
    }
    \subfigure[$\left<\cos \theta \right>(r)$]{
      \includegraphics[width=3.0in]{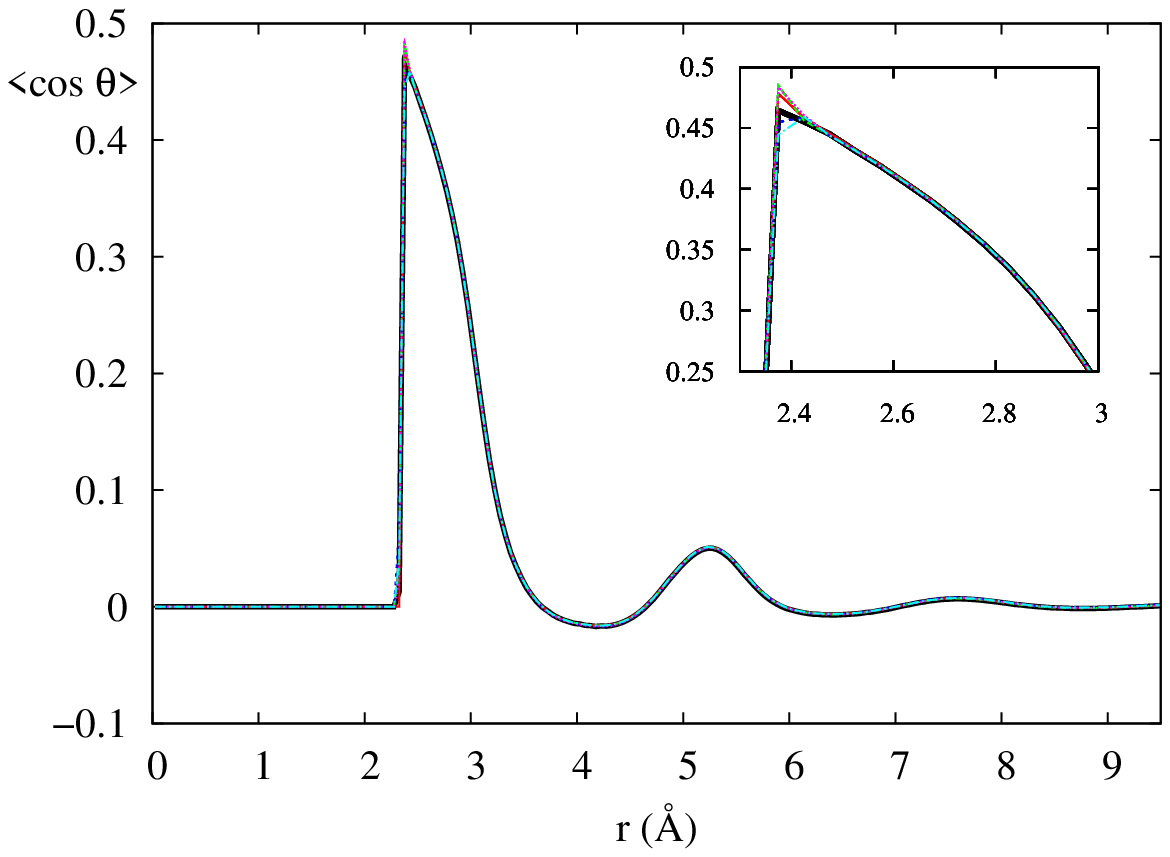}
    }
    \caption{$g(r)$ for each site-site pair of SPC/E water, as well as
      $\left< \cos \theta \right>(r)$.  All plots display the
      correlations determined via both Ewald summation (Full) and
      spherical truncation via LMF theory.  The smoothing
      length $\sigma$ ranges from 3.0~\AA\ to 6.0~\AA\ in all plots.  In
      the plot of $g_{\rm HH}(r)$, the curves for each $\sigma$ choice are
      displaced by 0.2 in order to emphasize that all plots shown
      contain multiple choices of \vs.  Insets in (a) and (d) focus on the region
      near the peak height, where small errors in the $\sigma = 3.0~\AA\ $ curves
      are just visible.}
    \label{fig:WaterCorrelations}
  \end{center}
\end{figure}

We have previously shown that
the SCA with a $\sigma$ of 4.5 \AA\ 
gives a highly accurate O-O pair
correlation function for SPC/E
water~\cite{RodgersWeeks.2008.Interplay-of-local-hydrogen-bonding-and-long-ranged-dipolar}.
In Fig.~\ref{fig:WaterCorrelations}(a-c), we show the pair correlation
functions for all site-site pairs using the SCA with $\sigma$ ranging from 3.0 \AA\
to 6.0 \AA.  In all instances, the $g(r)$ are in excellent agreement with
results of the full system determined using Ewald sums.  In the plot of
$g_{\rm HH}(r)$, the curves for each $\sigma$ choice are displaced by 0.2
in order to emphasize that all plots contain multiple choices of \vs\ while yielding essentially
the same correlation functions on the scale of the graph.

These data illustrate the important point that $\sigma$ is a consistency parameter
rather than an empirical fitting
parameter~\cite{ChenKaurWeeks.2004.Connecting-systems-with-short-and-long,RodgersWeeks.2008.Local-molecular-field-theory-for-the-treatment}. 
Thus the mean field averaging leading to LMF
theory become highly accurate for any choice of $\sigma$ greater than a state dependent
minimum value $\sigma_{\text{min}}$,
typically of order a characteristic nearest neighbor distance.
For SPC/E water $\sigma_{\text{min}}$ is about $3~\AA$, the radius of the
Lennard-Jones (LJ) core on the water molecule. Any smaller $ \sigma $ would
clearly yield a short-ranged system that does not accurately describe the
oxygen-hydrogen charge correlations on neighboring molecules that
compete with the LJ core repulsions in forming hydrogen bonds, and indeed poor
agreement is found at smaller $\sigma$.

We also carried out molecular dynamics simulations of bulk acetonitrile
at two very different states, a high density liquid very near liquid-vapor coexistence
at 298~K and a lower density system at 550~K. We used a six-site model
with flexible bonds developed by Nikitin and
Lyubartsev~\cite{NikitinLyubartsev.2007.New-six-site-acetonitrile-model-for-simulations-of-liquid}
in which intermolecular potential parameters have been optimized for
better consistency with experiments. In order to simulate at
appropriate bulk densities, an initial configuration of 864 molecules
in a cubic box is equilibrated for several hundred picoseconds (ps) in
the NPT ensemble using the Nose-Hoover thermostat and barostat until
the volume has equilibrated.  The low temperature system has a
simulation box length of 42.2~\AA. The dilute system at 550~K has a
density one-third that of 298~K and is further equilibrated in the NVT
ensemble for several hundred picoseconds.  The cutoffs of the
Lennard-Jones interactions were set to 15~\AA.  When the SCA is
employed, the cutoffs for \vs\ were 15~\AA\ for
$\sigma=4.5$~\AA, 21~\AA\ for $\sigma=6.5$~\AA, and 30~\AA\ for
$\sigma=8.5$~\AA.  When Ewald summation was employed as a benchmark,
the cutoff for the real space interactions was set to 15~\AA, and
$\alpha=0.26$~\AA$^{-1}$ with $\vect{k}_{\rm max}=(15,15,15)$.

\begin{figure}
  \begin{center}
    \subfigure[$g_{\rm YN-CT}(r)$]{
      \includegraphics[width=3.0in]{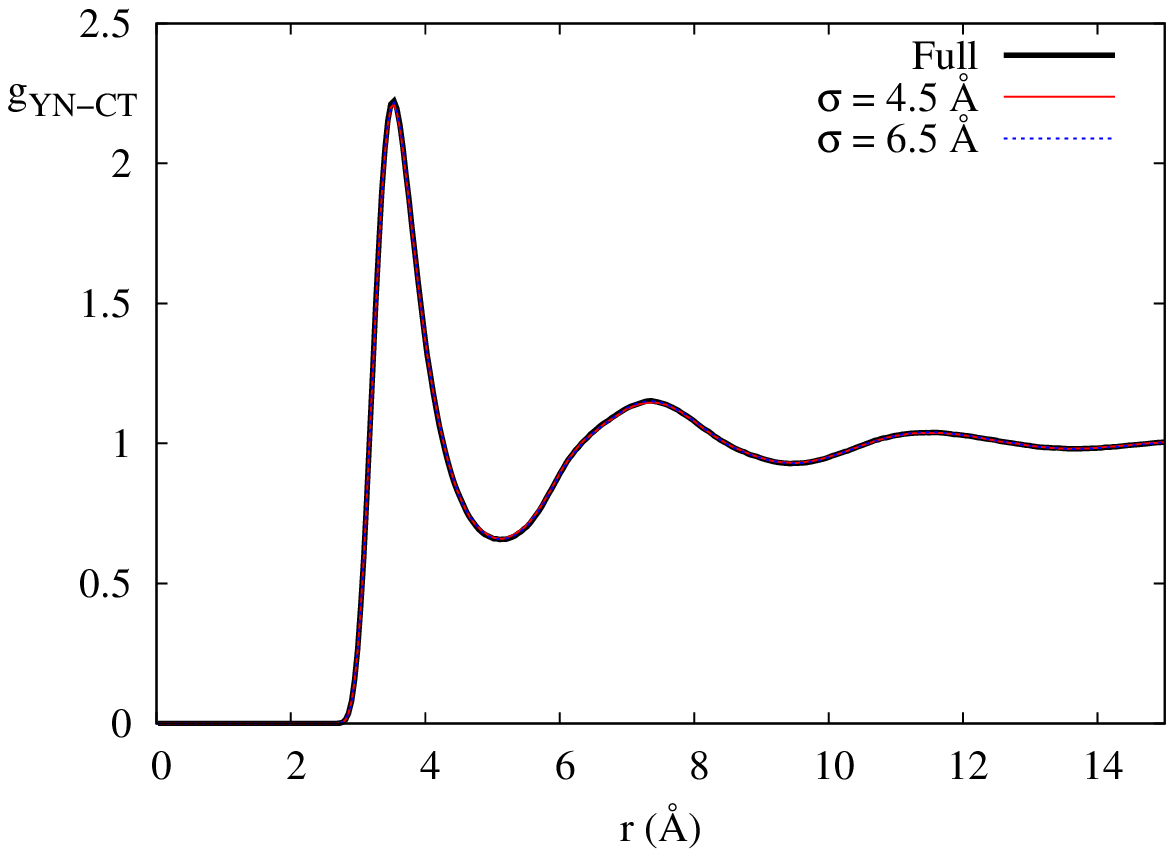}
    }
    \subfigure[$g_{\rm YN-HC}(r)$]{
      \includegraphics[width=3.0in]{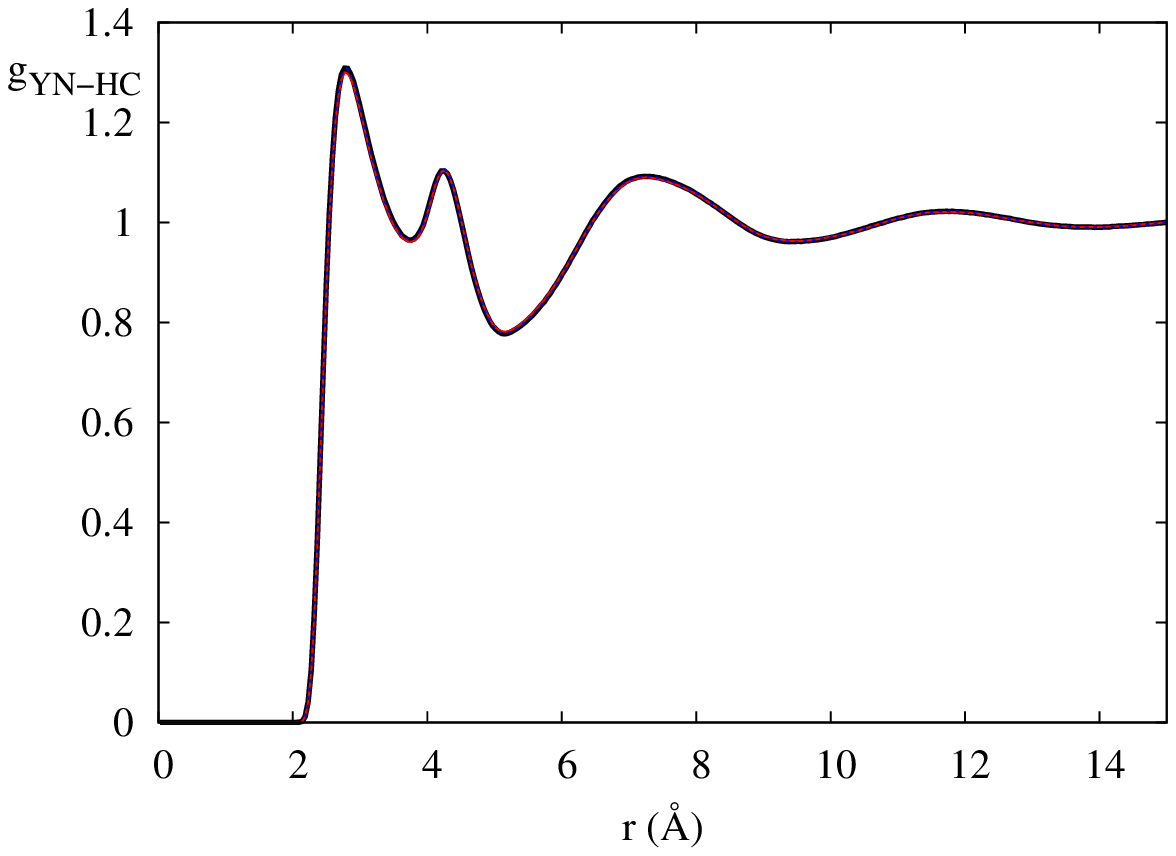}
    }
   \subfigure[$g_{\rm YN-YC}(r)$]{
      \includegraphics[width=3.0in]{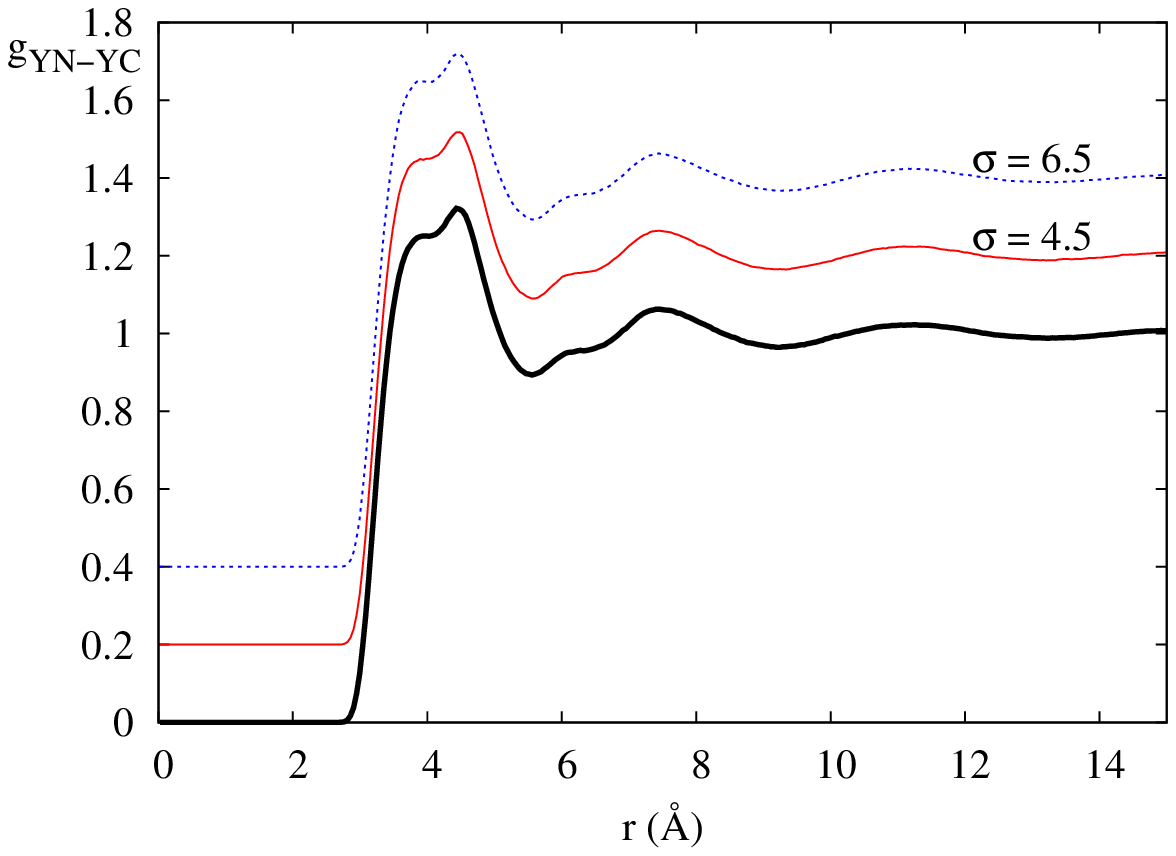}
    }
    \subfigure[$g_{\rm YN-YN}(r)$]{
      \includegraphics[width=3.0in]{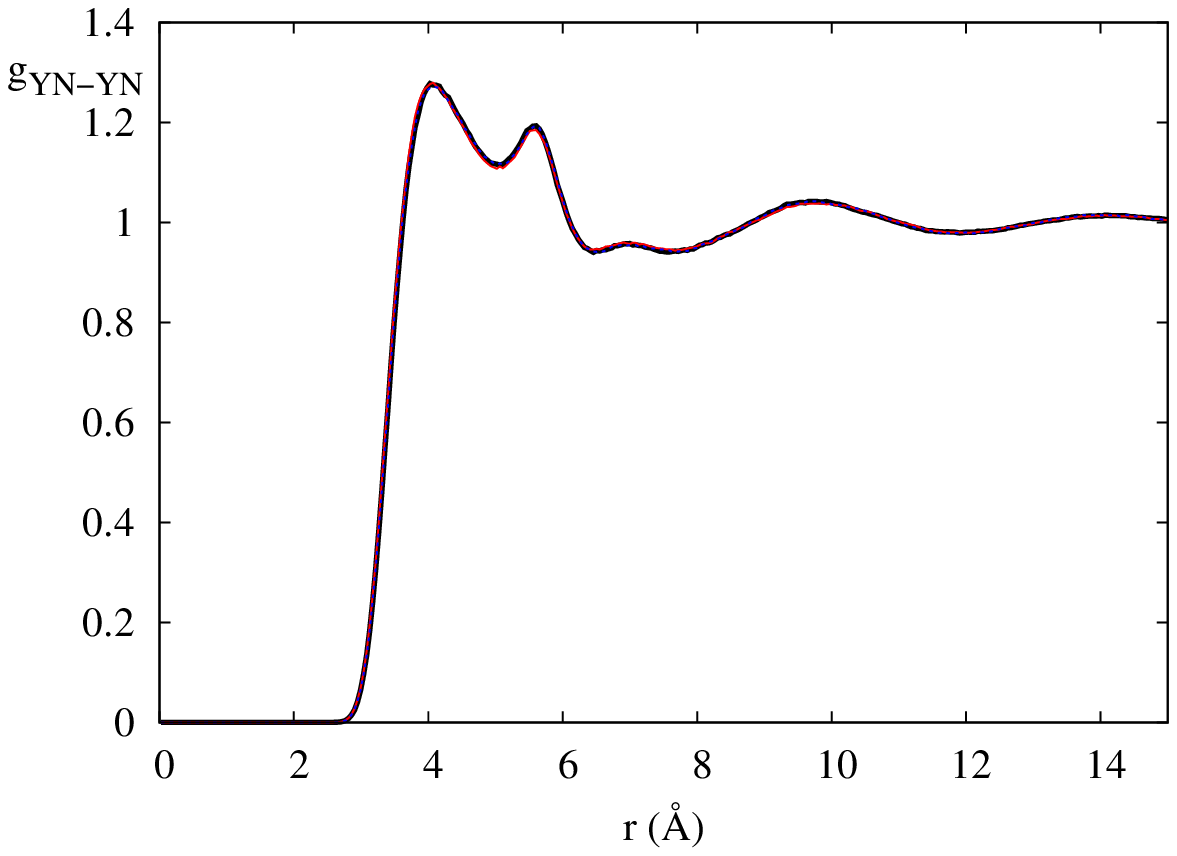}
    }
    \caption{Pair correlation functions for the nitrogen site
      YN on acetonitrile at 298~K as $\sigma$ varies from 4.5~\AA\ to
      6.5~\AA.  As before, the plot of $g_{\rm YN-YC}(r)$ displays the
      $g(r)$ displaced by 0.2 but all plots display results for Ewald
      summation (Full) and all choices of $\sigma$.}
    \label{fig:AcetonitrileCorrelationsT298}
  \end{center}
\end{figure}

\begin{figure}
  \begin{center}
    \subfigure[$g_{\rm YN-CT}(r)$]{
      \includegraphics[width=3.0in]{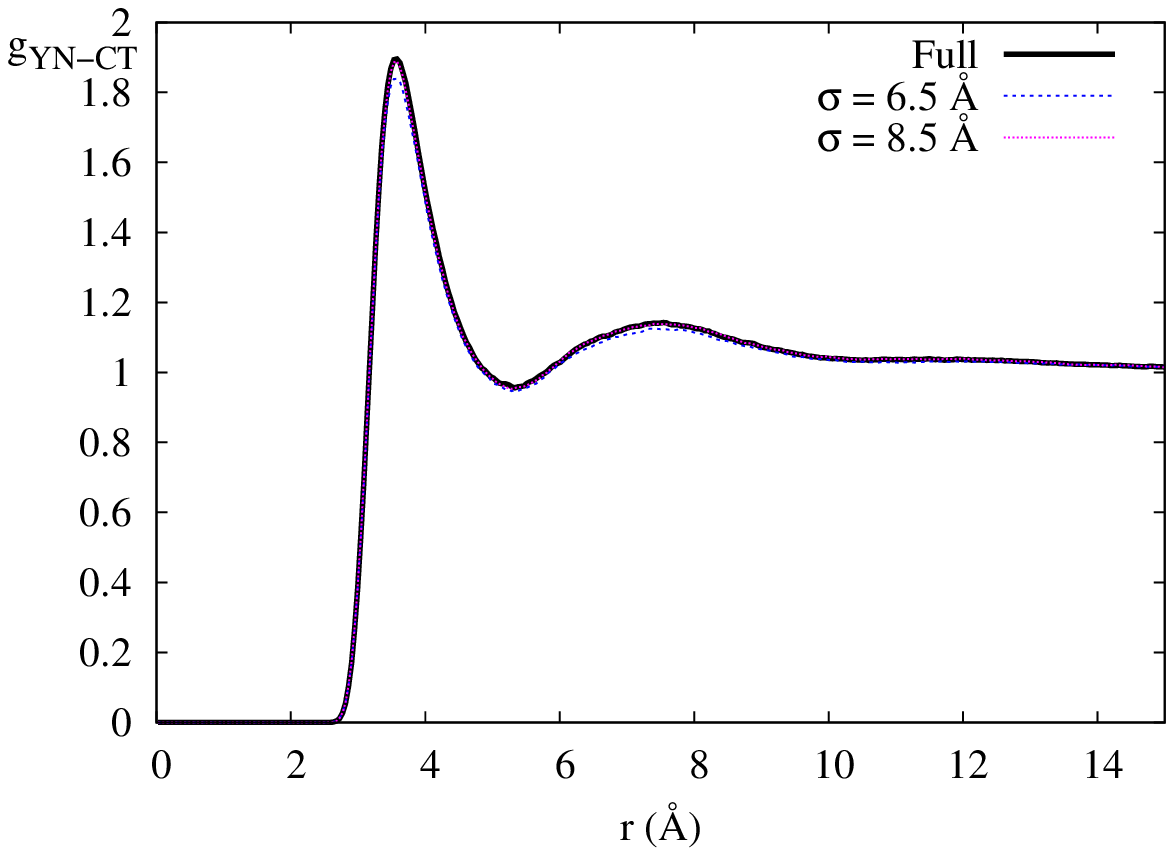}
    }
    \subfigure[$g_{\rm YN-HC}(r)$]{
      \includegraphics[width=3.0in]{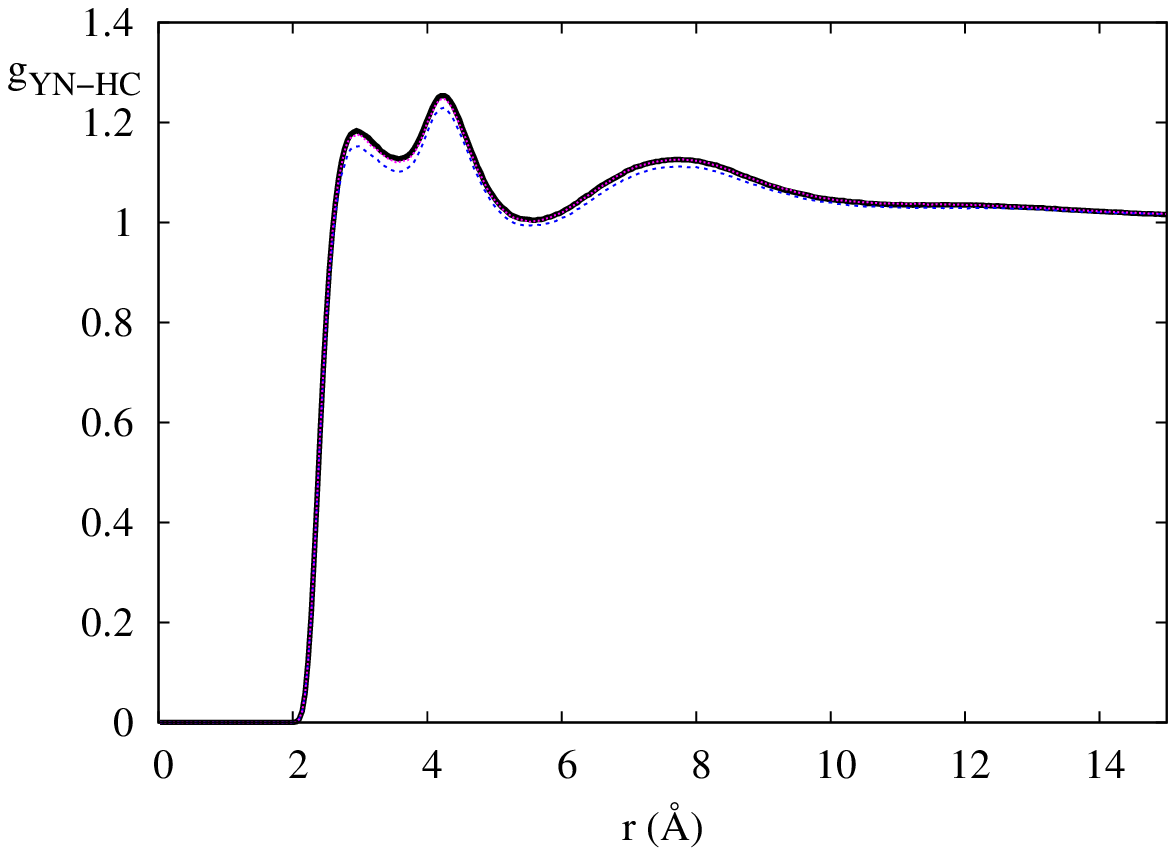}
    }
   \subfigure[$g_{\rm YN-YC}(r)$]{
      \includegraphics[width=3.0in]{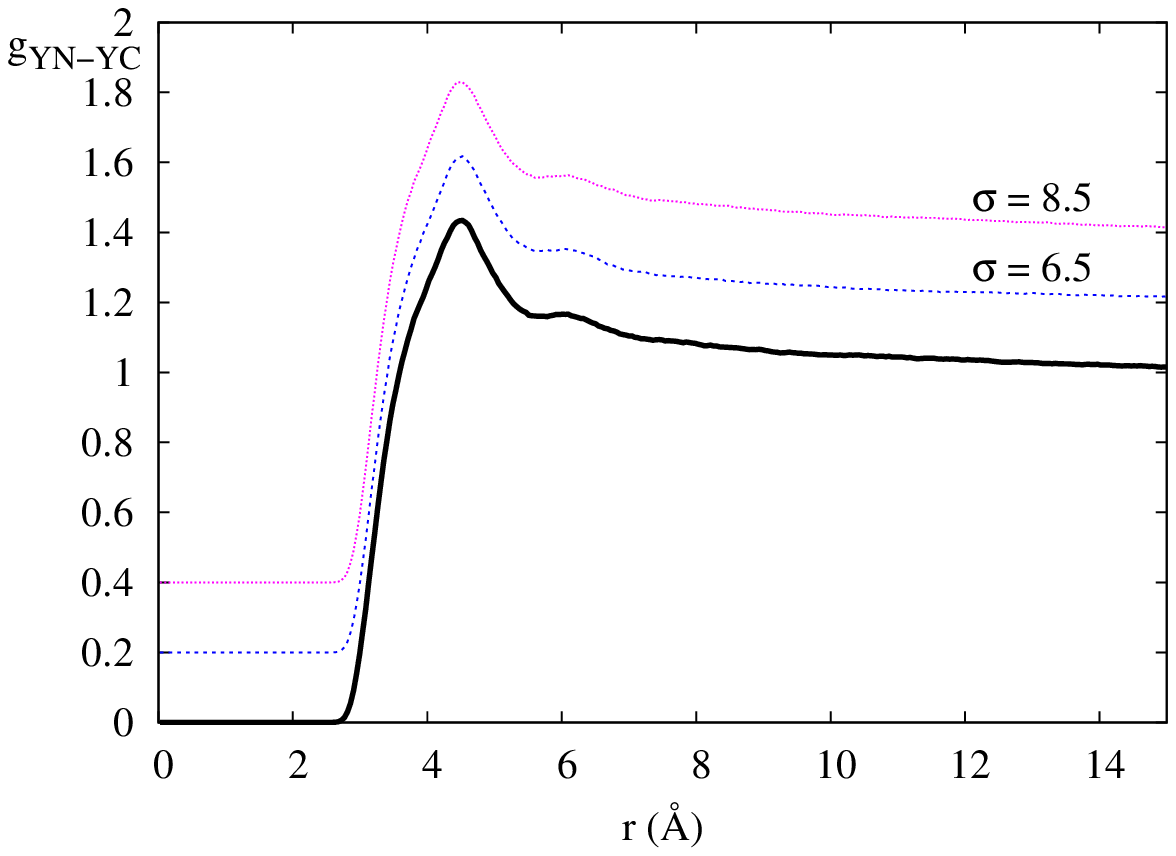}
    }
    \subfigure[$g_{\rm YN-YN}(r)$]{
      \includegraphics[width=3.0in]{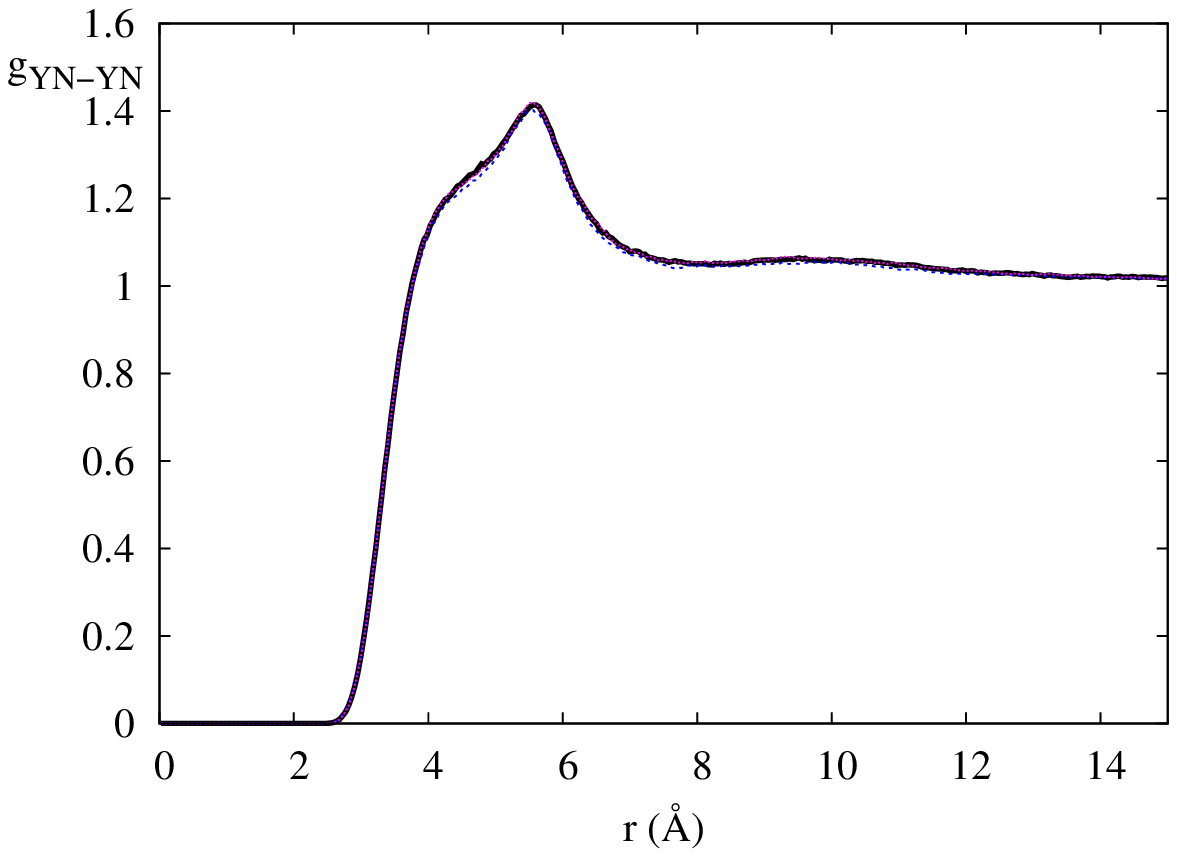}
    }
    \caption{Pair correlation functions for the nitrogen site YN on
      acetonitrile at 550~K as $\sigma$ varies from 6.5~\AA\ to
      8.5~\AA.  As before, the plot of $g_{\rm YN-YC}(r)$ displays the
      $g(r)$ displaced by 0.2 but all plots display results for Ewald
      summation (Full) and all choices of $\sigma$.}
    \label{fig:AcetonitrileCorrelationsT550}
  \end{center}
\end{figure}

Results for acetonitrile site-site pair correlation functions are shown in
Figs.~\ref{fig:AcetonitrileCorrelationsT298}
and~\ref{fig:AcetonitrileCorrelationsT550}.  These figures focus on
the pair correlations at
both 298~K and 550~K between a nitrogen site (YN) and all four molecular sites
on another molecule. 
The remaining six intermolecular site-site pair correlations are described
just as accurately, and are not
displayed for brevity.  For the high density room temperature system,
both $\sigma$ shown yield quite accurate results. 
Note that the $\sigma$ values are comparable to those used for water,
despite the greater size of the acetonitrile molecule. Very poor results (not shown)
were obtained with use of a too small $\sigma=2.5$~\AA\ as would be expected.

For the higher temperature, lower density system,
$\sigma=4.5$~\AA\ performs poorly and is not shown, $\sigma=6.5$~\AA\ is markedly
improved, and only the largest $\sigma$ of 8.5~\AA\ yields high
quality agreement with the Ewald simulation.  This is expected from
simple scaling arguments since the typical
nearest neighbor distance is larger; multiplying $\sigma_{\rm min}=4.5$~\AA\
for 298~K by the requisite increase in interparticle spacing at lower
densities yields 6.5~\AA.  The need for a somewhat larger
$\sigma_{\rm min}$ is likely a result of the increasing relevance of more
extended conformations of these molecules at lower densities and
higher temperatures.  Both the water and acetonitrile results show
that spherical truncations are quite good in bulk fluids, given a
sufficiently large truncation radius.  This is phenomenologically well
established in the literature.

The strong agreement of all the acetonitrile site-site correlation
functions, given a sufficiently large $\sigma$, suggests that the
angular correlations between these molecules are also accurate, for
otherwise many of the unusual functional forms would not be reproduced
with fidelity.  Thus we also examine angular correlations for both water and acetonitrile.

\begin{figure}
  \begin{center}
    \subfigure[$T=298$~K]{
      \includegraphics[width=3.0in]{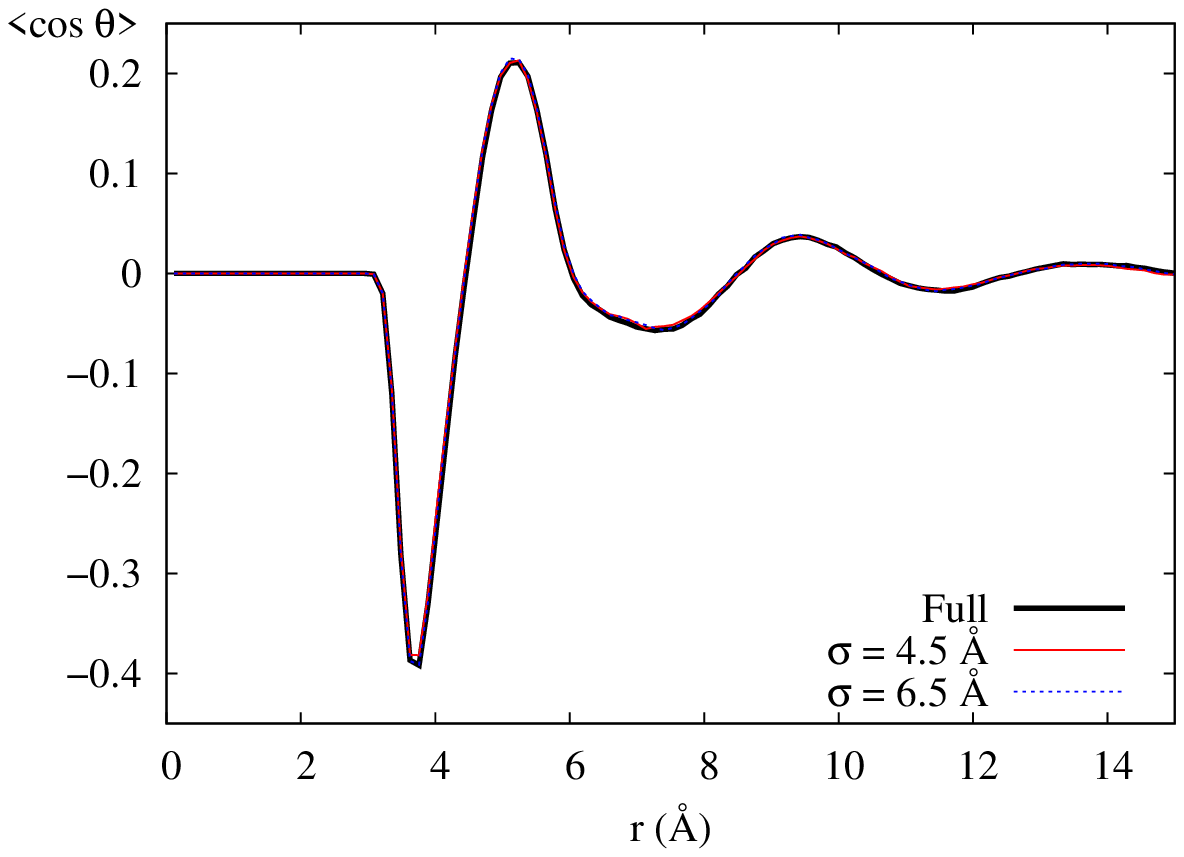}
    }
    \subfigure[$T=550$~K]{
      \includegraphics[width=3.0in]{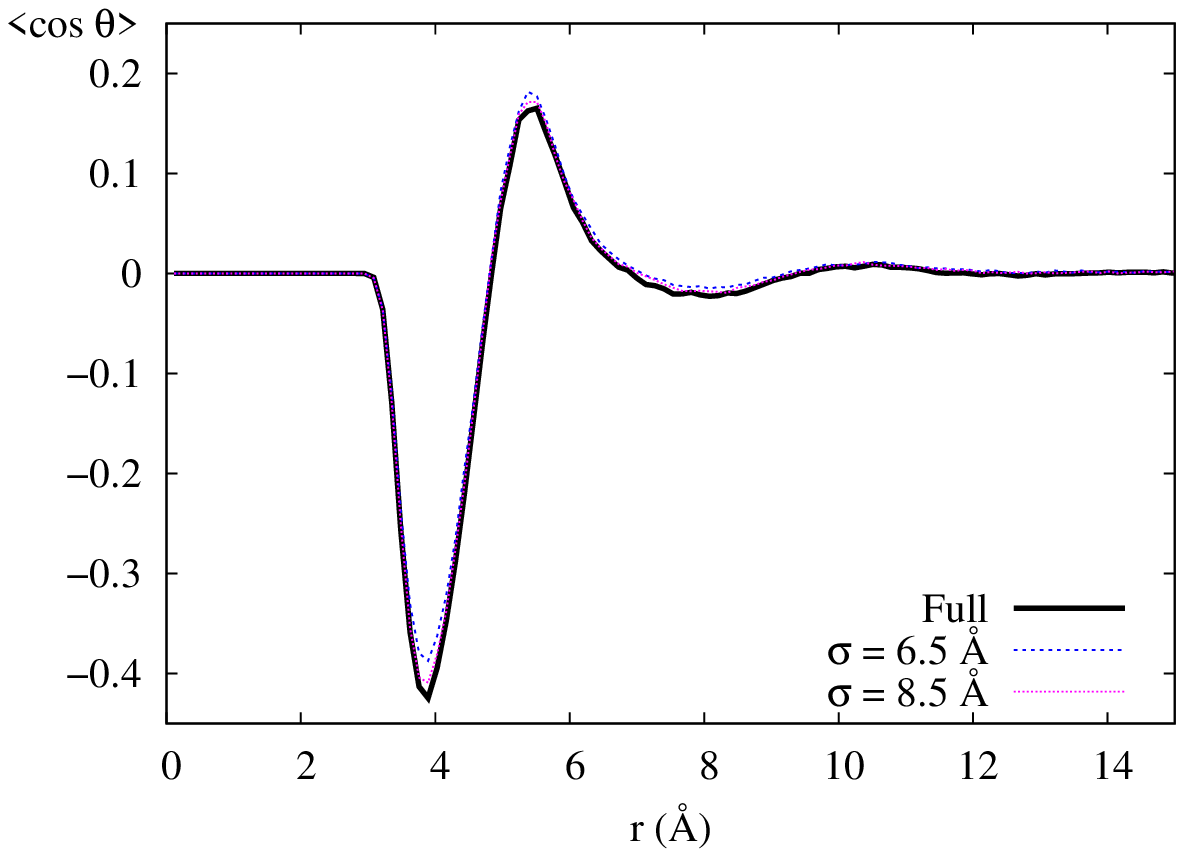}
    }
    \caption{Angular correlations represented by $\left< \cos \theta
      \right>(r)$ for acetonitrile at 298~K (left) and 550~K (right) as
      $\sigma$ varies.}
    \label{fig:AcetonitrileAngular}
  \end{center}
\end{figure}

Fig.~\ref{fig:WaterCorrelations}(d) shows the excellent agreement of
dipole-dipole correlations
in \spce\ water simulated using the SCA with those correlations in the
full Ewald simulations.
Here we plot the average $\cos \theta$ between water dipoles
as a function of separation distance $r$ between the centers of mass
of two water molecules.  Such good agreement is not a consequence of
the relatively compact nature of the water molecule.  Shown in
Fig.~\ref{fig:AcetonitrileAngular} are plots of $\left< \cos \theta
\right>(r)$ for the acetonitrile system at each temperature.  We again
find quite good agreement between the angular correlations in the full
Ewald system and our short-ranged systems.  This agreement follows the
trends found for the simpler site-site distribution functions, with
a larger $\sigma$ needed for the low density higher temperature state.

We believe that the excellent agreement of the
dipole-dipole correlations in these spherically truncated fluids is a
direct consequence of the general validity of LMF theory and the accuracy of
the strong coupling approximation in uniform environments, as we describe in the following
section.  Nezbeda has previously reported poor results for
dipole-dipole correlations in short-ranged systems where the
determined $g(r)$ were
accurate~\cite{Nezbeda.2005.Towards-a-Unified-View-of-Fluids}.  The
crux of the difficulties with his chosen cutoff scheme was defining
these cutoffs on a molecular basis, rather than a site basis.  This
leads to neglected potentials which actually rapidly vary near the
cutoff radius, counter to one of the important assumptions of \lmf\
theory as discussed in the next section.  Takahashi and
coworkers~\cite{TakahashiNarumiYasuoka.2010.Cutoff-radius-effect-of-the-isotropic-periodic}
studied the effect of cutoff radii in the isotropic periodic sum
approach on various properties of water and found for $\left< \cos
  \theta \right>(r)$ that deviations in this property were minimal and
equivalent for cutoff radii greater than 16~\AA.  This cutoff radius
for IPS can be compared to the cutoff radius of 13.5~\AA\ used for
$\sigma=6.0$~\AA\ in this paper.  We take their observed
``saturation'' in errors beyond a given cutoff radius as an indication
that their spherically-truncated potential satisfies the necessary
conditions for the validity of LMF theory.

\section{Local Molecular Field (LMF) Theory for Site-Site
  Molecules \label{sxn:LMFdiscussion}}

LMF theory for a general nonuniform system prescribes a mapping
from the system of interest where
all particles interact via their full intermolecular potentials in the
presence of an external field to a ``mimic system''  where particles interact via short-ranged
truncations of their intermolecular potentials but in the presence of a an effective or
restructured field. The restructured field accounts for the effects of the
long-ranged components of the intermolecular interactions using a mean field
average.  Far from being a simplistic mean-field ansatz,
LMF theory has been shown to be strongly rooted in statistical
mechanical theory, and based on physically-motivated approximations
that are well-justified for dense fluid systems.

LMF theory for charged 
systems takes a particularly simple form when charges alone
are treated via \lmf\ theory using a single $\sigma$, where the results can be
exactly re-expressed in terms of the total charge density and a
restructured electrostatic potential.
Based on the splitting of $1/r$ defined in equation~(\ref{eq:CoulombSplit}),
each pair potential $u_{\alpha \xi}(r)$ in equation~(\ref{eq:uniformU}) may
be decomposed as
\begin{equation}
  u_{\alpha \xi}(r) = u_{0,\alpha \xi}(r) + \frac{q_\alpha q_\xi}{\epsilon} \vl,
\end{equation}
where $u_{0,\alpha \xi}(r)$ contains all the non-electrostatic
Lennard-Jones-like pair interactions as well as a \vs\ contribution
appropriately scaled by charge and the dielectric constant $\epsilon$.
The crucial feature of these two potentials \vs\ and \vl\ for the
validity of LMF theory is that $\sigma$ is chosen so that \vs\ contains all relevant strong
Coulomb forces between nearest neighbors and that \vl\ is
consequently slowly-varying
over the range of strongest correlations between those neighbors.  A more
careful statement of the relevant approximations may be found in
Appendix~\ref{app:SimpleYBGandLMF}.

Previous work
focused on nonuniformity such as confining walls, using the 
Coulomb LMF equation ~\cite{RodgersWeeks.2008.Interplay-of-local-hydrogen-bonding-and-long-ranged-dipolar,RodgersWeeks.2008.Local-molecular-field-theory-for-the-treatment}
\begin{equation}
  \Vr(\vect{r}) = \mathcal{V}_0(\vect{r}) + \frac{1}{\epsilon} \int
  d\vect{r}^\prime \, \rho^q_{R,\tot}(\vect{r}^\prime) v_1
  \left(\vdiff{r}{r^\prime}\right),
 \label{eq:LMFgeneralcoulombNonuniform}
\end{equation}
where \Vs\ results from the convolution of the fixed charge density
with \vs, and $\rho^q_{R,\tot}$ includes both the fixed and mobile
charge densities. Note that \Vr\ and $\rho^q_{R,\tot}$ are implicitly
functionals of one another, so this is a self-consistent equation.

Since \vl\ is the electrostatic potential arising from a Gaussian charge
density with width $\sigma$, equation~(\ref{eq:LMFgeneralcoulombNonuniform})
suggests that the
restructured external potential \Vr\ may be understood as an
electrostatic potential containing the full impact of fixed charges
and the Gaussian-smoothed impact of mobile charges.
In Appendix~\ref{app:SimpleYBGandLMF}, we present a derivation of the LMF
equation for site-site molecular models as used in previous papers.
Equation~(\ref{eq:LMFgeneralcoulombNonuniform}) is identical to that for mixtures of charged
species~\cite{RodgersWeeks.2008.Local-molecular-field-theory-for-the-treatment},
and a derivation for small site-site molecular models requires only
one further approximation, requiring that intramolecular correlations
are well represented by the mimic system, a seemingly very reasonable
requirement.  The solution of equation~(\ref{eq:LMFgeneralcoulombNonuniform}) has been shown to
yield accurate structure for both ionic
solutions~\cite{ChenWeeks.2006.Local-molecular-field-theory-for-effective,RodgersKaurChen.2006.Attraction-between-like-charged-walls:-Short-ranged}
and molecular
water~\cite{RodgersWeeks.2008.Interplay-of-local-hydrogen-bonding-and-long-ranged-dipolar}
in nonuniform systems, and a simple linear response method for solving the
above equation has been
derived~\cite{HuWeeks.2010.Efficient-Solutions-of-Self-Consistent-Mean-Field},
leading to fast and computationally efficient solutions of the LMF equation.

Site-site pair correlations in bulk fluids may be simply related to those arising from
fixing a given site at the origin, thus allowing us to describe
structure in uniform fluids from the nonuniform perspective
of LMF theory in equation~(\ref{eq:LMFgeneralcoulombNonuniform}). 
As such, for bulk molecular fluids
with spherically symmetric site-site interactions as considered here,
we would expect that the general LMF
equation~(\ref{eq:LMFgeneralcoulombNonuniform}) in this case could then be written as
\begin{equation} {\mathcal{V}_{R|\eta}}(r) =
  \frac{q_\eta}{\epsilon}\vs+ \frac{1}{\epsilon} \int d\vect{r}^\prime
  \, \rho_{R,\tot}^{q}({r^{\prime} |{\mathbf 0}}) v_1
  \left(\vdiff{r}{r^\prime}\right),
  \label{eq:LMFgeneralcoulombInText}
\end{equation}
with $\eta$ being the site fixed at the origin as indicated by the
conditional notation $| \eta $ on the left side of equation (\ref{eq:LMFgeneralcoulombInText}).
In analogy with
equation~(\ref{eq:LMFgeneralcoulombNonuniform}), the first term is the
short-ranged potential due to the only fixed charge in the system,
the charge from site $\eta$ at the origin.  The charge density
$\rho_{R,\tot}^q(r^\prime | {\mathbf 0})$ is the total charge density
in the nonuniform mimic system with $|{\mathbf 0}$
again indicating the fixed site.  In the case of these small site-site
molecules, this total charge density may be decomposed as
\begin{itemize}
\item the intramolecular charge density arranged around the molecular
site $\eta$ fixed at the origin including the site $\eta$, denoted
 $\varrho^{q}_{R,M|\eta}({r}|{\mathbf 0})$, and
\item the intermolecular charge density from other unconstrained mobile
molecules induced by $\eta$, denoted $\rho^{q}_{R}(r|{\mathbf 0})$.
\end{itemize}
While only $\eta$ contributes to the \Vs\ in this equation, the
intramolecular sites attached to the site $\eta$ fixed at the origin
contribute directly to the total charge density $\rho_{R,\tot}^{q}$
and also implicitly but strongly impact the form of the intermolecular
charge density $\rho^q_R$ based
on their inclusion in the simulation of the mimic system.

While the discussion above should make the form of
equation~(\ref{eq:LMFgeneralcoulombInText}) quite plausible, two further
approximations are needed in addition to the three stated in
Appendix~\ref{app:SimpleYBGandLMF} to carefully separate effects of intra-
and intermolecular charges in this equation
and to assess its accuracy for site-site molecular models.
Again, we employ the exact relationship between the pair distribution functions
in a uniform fluid and the conditional singlet density profile due to
a site fixed at the origin. 

The YBG hierarchy for site-site molecular
systems with one site of one molecule fixed at the origin is derived
in Appendix~\ref{app:YBG}.  The derivation is quite interesting technically,
since we use an external field to localize only a particular molecular site at the origin rather
than to represent an entire fixed molecule, as is usually done.
Moreover,  we derive first the YBG hierarchy for correlation functions
between specific molecules
rather than the usual generic correlation functions used in standard treatments.
These features allows us to more easily disentangle contributions
from intra- and intermolecular correlation functions. Using this new YBG hierarchy, the
derivation of LMF theory for a uniform fluid of site-site molecules
then follows the traditional route, while requiring two new
but very plausible approximations related to intramolecular correlations, as shown in
Appendix~\ref{app:LMFbulk}. This provides a rigorous derivation of
equation~(\ref{eq:LMFgeneralcoulombInText}).

Towards the goal of understanding the accuracy of SCA truncations
in uniform fluids,
we rewrite equation~(\ref{eq:LMFgeneralcoulombInText}) in a way that
focuses on the \emph{long-ranged} contributions to \Vr, which we call \Vrl:
\begin{equation} {\mathcal{V}_{R1|\eta}}(r) \equiv
  \mathcal{V}_{R|\eta}(r) - \frac{q_\eta}{\epsilon}\vs =
  \frac{1}{\epsilon} \int d\vect{r}^\prime \,
  \rho_{R,\tot}^{q}({r^{\prime} |{\mathbf 0}}) v_1
  \left(\vdiff{r}{r^\prime}\right).
  \label{eq:LMFgeneralVRlongcoulombInText}
\end{equation}
If $\mathcal{V}_{R1|\eta} \approx 0$
then simulating with spherical truncations alone
as in the SCA will give very accurate results.

This $\mathcal{V}_{R1|\eta}$ defined in
equation~(\ref{eq:LMFgeneralVRlongcoulombInText}) is seen to be the
Gaussian-smoothed electrostatic potential arising from
the total charge density in the fluid induced by the charge
from fixed site $\eta$. 
This total charge density includes the single molecule charge
distribution $\varrho^{q}_{R,M|\eta}({r}|{\mathbf 0})$ as well as contributions
from other fully mobile molecules.
As discussed in
Refs.~\cite{RodgersWeeks.2008.Local-molecular-field-theory-for-the-treatment},~\cite{RodgersWeeks.2008.Interplay-of-local-hydrogen-bonding-and-long-ranged-dipolar},
and~\cite{ChenWeeks.2006.Local-molecular-field-theory-for-effective},
the restructured electrostatic potential $\Vrl(\vect{r})$ induced by a
general fixed charge distribution $\rho^{q}_{\rm
  ext}(\vect{r}^\prime)$ satisfies the single Coulomb LMF equation
given by the convolution of \vl\ with
$\rho^q_{R,\tot}(\vect{r}^\prime)$, including contributions from both
fixed and mobile charges, so equation~(\ref{eq:LMFgeneralVRlongcoulombInText})
has exactly the form that would be expected.

\section{Success of SCA Explained\label{sxn:SCAsuccess}}
We specifically explore the meaning and consequences of the LMF
equation for SPC/E water, both because it has fewer sites than
acetonitrile and also because it has a fixed geometry, thereby
allowing for analytical determination of
$\varrho^{q}_{R,M|\eta}({r}|{\mathbf 0})$ without simulation
and independent of perturbations from other mobile molecules. For
either hydrogen site,
\begin{equation}
  \varrho^{q}_{R,M|H}(r|{\mathbf 0}) = q_H \delta\left( \vect{r} \right) + q_O \frac{\delta \left( r - r_{\rm OH} \right)}{4 \pi r_{\rm OH}^2} + q_H \frac{\delta \left( r - r_{\rm HH}\right)}{4 \pi r_{\rm HH}^2},
  \label{eq:Hsite}
\end{equation}
and for the oxygen site,
\begin{equation}
  \varrho^{q}_{R,M|O}(r|{\mathbf 0}) = q_O \delta\left( \vect{r} \right) + 2 q_H \frac{\delta \left( r - r_{\rm OH} \right)}{4 \pi r_{\rm OH}^2},
  \label{eq:Osite}
\end{equation}
where the charge densities have been spherically averaged about the
site fixed at the origin.  
Separating out the contribution of these intramolecular charge densities, 
the total ${\mathcal{V}_{R1|\eta}}$ in
equation~(\ref{eq:LMFgeneralVRlongcoulombInText}) may be decomposed into
intramolecular and intermolecular contributions as
\begin{equation} 
  {\mathcal{V}_{R1|\eta}}(r) =
  \frac{1}{\epsilon} \int d\vect{r}^\prime \,
  \varrho_{R,M|\eta}^{q}({r^{\prime} |{\mathbf 0}}) v_1
  \left(\vdiff{r}{r^\prime}\right) + 
  \frac{1}{\epsilon} \int d\vect{r}^\prime \,
  \rho_{R}^{q}({r^{\prime} |{\mathbf 0}}) v_1
  \left(\vdiff{r}{r^\prime}\right),
  \label{eq:LMFgeneralIntraInterInText}
\end{equation}
where the first term corresponds to the long-ranged interactions due
to sites within the molecule with site $\eta$ at the origin.
 
The success of SCA shown in Section~\ref{sxn:SCAresults} suggests that
$\mathcal{V}_{R1|\eta} \approx 0$ is a well-founded approximation.
Before utilizing the analytical charge densities above, we first explore an
alternate formulation of LMF theory for site-site molecules which
might seem initially fruitful.  Theoretical treatments of molecular
models often involve fixing a given molecular orientation and
considering the fluid response to this configuration.  Based on the
splitting of $1/r$, the majority of the strong electrostatic potential energy
and force will be included in the \vs\ used as a pair potential in
SCA simulations.  However, for any one orientation of a water molecule,
the combined forces due to the \vl\ on other oxygen and hydrogen sites
will not be negligible, even though they are slowly-varying on the scale of
\sig\ for each individual site. 

\begin{figure}[t]
  \begin{center}
    \includegraphics[width=3.0in,angle=-90]{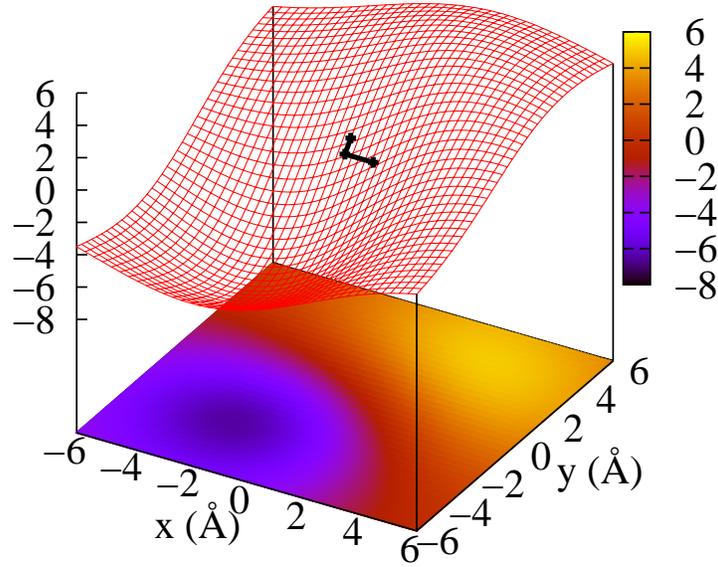}
    \caption{The long-ranged potential in the z=0 plane resulting from a
      fixed orientation of a water molecule with $\vect{r}_{\rm
        O}=(0,0,0)$, $\vect{r}_{\rm H1}=(1,0,0)$, and $\vect{r}_{\rm
        H2}=(-0.334,0.943,0)$ and \sig\ chosen as 4.5~\AA.  The potential
      is displayed in units of $k_BT / e_0$ in order to aid in gauging
      the magnitude of this potential relative to thermal fluctuations.
      The chosen orientation of the water molecule is shown with solid
      black lines and points.}
    \label{fig:V1WaterOrient}
  \end{center}
\end{figure}

As one example of this, the
long-ranged electrostatic potential arising from a fixed orientation of a
water molecule with O at the origin and the $1/r$ interactions replaced by
\vl\ with $\sigma=4.5$~\AA\ is shown in
Fig.~\ref{fig:V1WaterOrient}.  Based on this single snapshot of the
\vl\ contributions due to intramolecular sites, neglect of these
long-ranged forces in the SCA would seem an ill-conceived approximation, and we
might suppose that a \Vr\ depending on both intermolecular distance
and relative molecular orientations would be required.  However,
looking at an individual orientation of the water molecule for long-ranged
interactions fixes all three intramolecular charges and would
be expected to generate a very different and larger density response
than the single fixed molecular charge at
the origin needed to determine radially-symmetric site-site
correlation functions, as the \lmf\
equation~(\ref{eq:LMFgeneralIntraInterInText})
and Appendices~\ref{app:YBG} and~\ref{app:LMFbulk}
show.

\begin{figure}[b]
  \begin{center}
    \subfigure[Large Scale]{
      \includegraphics[width=3.0in]{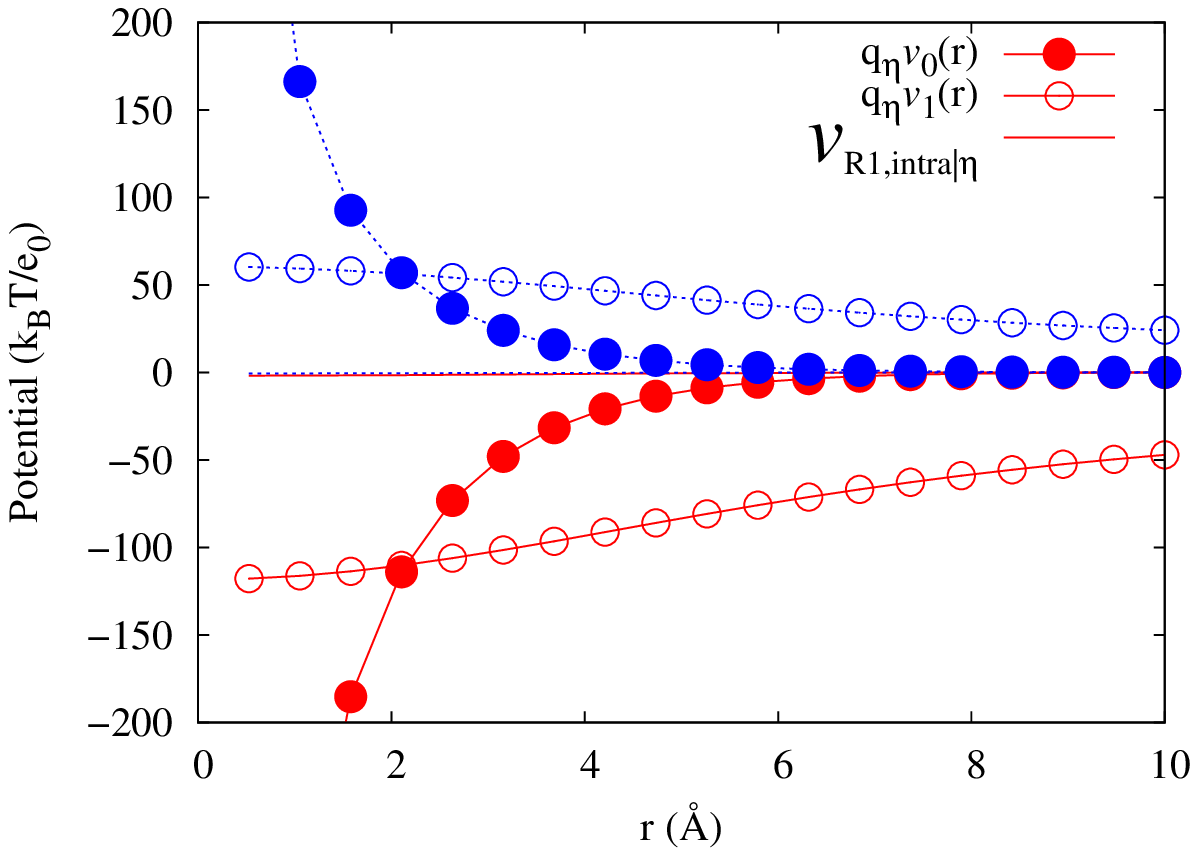}
    }
    \subfigure[Focus on $\mathcal{V}_{R1,{\rm intra}|\eta}$]{
      \includegraphics[width=3.0in]{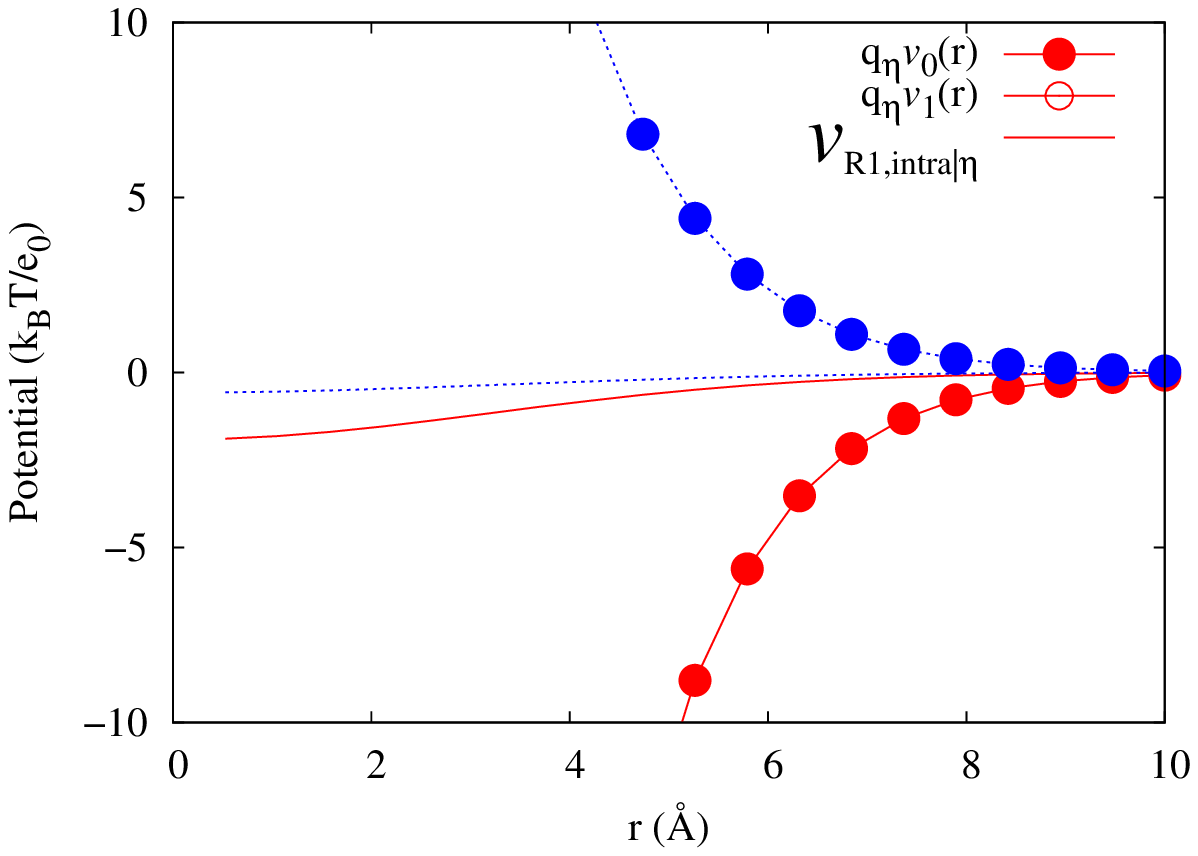}
    }
    \caption{Comparison of $\mathcal{V}_{R1,{\rm intra}|\eta}(r)$ to
      relevant potentials due solely to the site $\eta$ fixed at the
      origin, whether it be oxygen (red) or hydrogen (blue).  This
      electrostatic potential due to the whole neutral molecule is substantially
      smaller than both the short-ranged and long-ranged components of
      $1/r$ due to the individual site fixed at the origin.}
    \label{fig:V1WaterSpherical}
  \end{center}
\end{figure}

The first term in equation~(\ref{eq:LMFgeneralIntraInterInText}) may be
determined analytically for SPC/E water, and this is the first crucial step in
understanding why the full $\mathcal{V}_{R1|\eta}$ will be small
to good approximation in uniform systems.  As shown
in Fig.~\ref{fig:V1WaterSpherical}, the spherically symmetric
long-ranged potential from the first term, which we shall term $\mathcal{V}_{R1,{\rm
    intra}|\eta}(r)$ is indeed much more slowly-varying than the
orientation-dependent potential shown in Fig.~\ref{fig:V1WaterOrient}.
In this figure, we compare $\mathcal{V}_{R1,{\rm intra}|\eta}(r)$ to
both the \vs\ due to either O or H fixed at the origin as well as the
\vl\ due simply to that site $\eta$.

From these plots, we see that $\mathcal{V}_{R1,{\rm intra}|\eta}$ is
substantially smaller in magnitude than either the electrostatic
potential arising from a specific water molecular orientation or the
potential due simply to the charge at the site $\eta$ we have fixed at the origin.
Therefore, we infer that the spherical truncations prescribed by \lmf\
theory and the associated mean-field averaging of long-ranged
interactions will actually be even more effective in bulk molecular
simulations than in a corresponding simulation
of an ionic system with charges not bound
into neutral molecules.  Again we emphasize that this spherical
averaging is not an unfounded approximation, but that it arises
rigorously from the statistical mechanics of molecular models interacting
via site-site potentials.

The \emph{total} electrostatic potential arising from the
spherically-averaged intramolecular charge density will be exactly
zero for all $r > l_{\rm OH}$ if oxygen is fixed at the origin or for
all $r > l_{\rm HH}$ if hydrogen is fixed at the origin.  Thus
it might seem counterintuitive that the corresponding
$\mathcal{V}_{R1,{\rm intra}|\eta}$ is small but non-vanishing
beyond this distance. However,
the distinct treatments of the short-ranged and long-ranged parts of
$1/r$ using Gaussian convolutions in LMF theory
require just such a nonzero potential.  All the
short-ranged parts of $1/r$ are treated explicitly via \vs\ positioned
around each site in the water molecule in order to represent local
correlations; the capture of these local correlations in the SCA
simulation is crucial.  In
tandem, only the long-ranged components \vl\ are spherically averaged
about the fixed site $\eta$ in LMF theory, leading to a \emph{non-zero} but
slowly-varying and small magnitude potential $\mathcal{V}_{R1,{\rm
intra}|\eta}(r)$ outside the total potential cutoff. 
For the correlations between molecules, the
need for non-zero short-ranged site-site $v_{0}$ terms
seems quite natural; the need for similar short-ranged
terms also holds for the far-less intuitively-obvious splitting of the
(exactly zero) electrostatic potential between two charged
plates~\cite{ChenWeeks.2006.Local-molecular-field-theory-for-effective}.

\begin{figure}
  \begin{center}
    \includegraphics[width=3.0in]{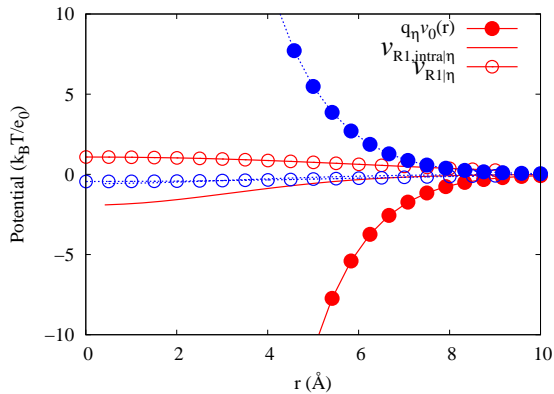}
    \caption{Estimation of $\mathcal{V}_{R1|\eta}$ based on charge
      densities from the simulations conducted using Ewald summation.}
    \label{fig:VR1ewald}
  \end{center}
\end{figure}

As demonstrated in Fig.~\ref{fig:V1WaterSpherical},
$\mathcal{V}_{R1,{\rm intra}|\eta}(r)$ is quite small and
slowly-varying for SPC/E water.  However, while this may make the
approximation that the total $\mathcal{V}_{R1|\eta}(r)\approx 0$ 
in equation~(\ref{eq:LMFgeneralIntraInterInText}) plausible, it
certainly does not guarantee it. Therefore, we also estimate
$\mathcal{V}_{R1|\eta}(r)$ by directly inserting the charge density
resulting from the Ewald simulation into the LMF equation.  The sole
care we take is in enforcing overall charge neutrality at the cutoff
radius for the potential, as this is also the furthest radius at which
$g(r)$ is calculated.  As seen in Fig.~\ref{fig:VR1ewald}, these
potentials are also quite slowly-varying, lending strong credence to
the approximation $\mathcal{V}_{R1|\eta}\approx 0$ in determining
structure.

This approach for determining $\mathcal{V}_{R1|\eta}(r)$ requires a
full Ewald simulation, contrary to the general philosophy of LMF theory, which
seeks to use simulations only in the mimic system. Thus strictly speaking
we should self-consistently solve for $\mathcal{V}_{R1|\eta}$
based on charge densities from the short-ranged mimic system using the
linear-response treatment developed in
Ref.~\cite{HuWeeks.2010.Efficient-Solutions-of-Self-Consistent-Mean-Field}.
But previous work has shown that the full LMF theory gives excellent agreement
with the results of Ewald simulations for water even in nonuniform
environments, so this Ewald determination should
be very accurate.
Furthermore, care should be exercised with the
$k=0$ component of any
charge density used in the LMF
equation~\cite{ChenKaurWeeks.2004.Connecting-systems-with-short-and-long,Rodgers.2008.Statistical-Mechanical-Theory-for-and-Simulations-of-Charged}.
However, charge densities obtained via Ewald summation exhibit
exponential screening and strictly
enforce overall neutrality, thus easing the need for great caution in the
treatment of the $k=0$ component. 

This simple estimate based on
the Ewald charge density certainly suffices to demonstrate
that $\mathcal{V}_{R1|\eta}(r)$ is small and slowly-varying in
this case, and provides strong justification for the accuracy of the SCA.  In general
we expect that quick estimates of \Vr\ using Ewald charge densities
when such simulations
are computationally practical will be very
useful in obtaining an accurate initial estimate of the final
self-consistent \Vr, and one that will be almost certainly in the linear regime where
the method of Ref.~\cite{HuWeeks.2010.Efficient-Solutions-of-Self-Consistent-Mean-Field}
will be especially easy to use.

However, for these bulk fluids an accurate \Vrl\ is neither necessary for
determining the structure to the accuracy shown here nor for determining the
thermodynamics of the fluid as shown in
Ref.~\cite{RodgersWeeks.2009.Accurate-thermodynamics-for-short-ranged-truncations-of-Coulomb}.
Provided that a sufficiently large $\sigma$ is chosen, simple
spherical truncations in simulations coupled with thermodynamic
perturbation theory yield accurate structure, energies, and pressures.
In the case of SPC/E water, structure might indicate that any $\sigma
\ge 3.0$~\AA\ is sufficiently large, but thermodynamics via
perturbation theory showed that $\sigma \ge 4.0$~\AA\ is
required~\cite{RodgersWeeks.2009.Accurate-thermodynamics-for-short-ranged-truncations-of-Coulomb}.

In general, the choice of a sufficiently large $\sigma$ is crucial for the accuracy of LMF theory.
For the acetonitrile system at the higher temperature and
lower density, inclusion of a self-consistent $\mathcal{V}_{R1}$ with
$\sigma=4.5$~\AA\ gives a poor description of the structure of the
acetonitrile system.  However, since our simple scaling analysis
suggests that $\sigma_{\rm min} \approx 6.5$~\AA, we do not expect LMF
theory with the smaller $\sigma$ to be able to correct this structure.  For the acetonitrile
systems at low and high temperature, just as for the water system at
ambient temperatures, a sufficiently large $\sigma$ yielded accurate
results simply via SCA.  Furthermore, the acetonitrile results already
demonstrate that $\sigma$ does not need to be on the scale of the
entire molecule but rather on the scale of nearest neighbor
correlations, as is expected from derivations of LMF  theory.
In Appendix~\ref{app:charmm}, we discuss LMF theory for
CHARMM-like molecules in order to better state the necessary
conditions for choice of $\sigma$ in much larger molecules.

\section{Conclusions \label{sxn:conc}}

In this paper, we have demonstrated the accurate results possible
using spherical truncations of $1/r$ interactions in simulations of uniform fluids.
We show that these spherical truncations yield not
only highly accurate pair correlation functions but also highly
accurate dipole-dipole correlation functions.
This good performance in bulk simulations of pair correlation functions was known; however, a solid
theoretical justification for the use of such spherical truncations in
molecular systems has been lacking.  In this paper, we present just
such a theoretical backing -- local molecular field theory.  The
derivations relevant to LMF theory for a variety of site-site
molecular models are presented in appendices and the
main paper focuses on understanding the accuracy of these spherical
truncations both phenomenologically and quantitatively using LMF
theory. LMF theory provides a general conceptual framework that helps
us understand why spherical truncations generally work so well in uniform systems
and also provides the essential corrections needed in most nonuniform environments.


\section*{Acknowledgements}

This work was supported by the National Science Foundation
(grants CHE0628178 and CHE0848574).
Z.H. acknowledges support from the startup funds provided
by Jilin University and The State Key Laboratory of Supramolecular Structure and Materials.
We are grateful to Rick Remsing and Shule Liu for helpful remarks.


\begin{thebibliography}{10}

\bibitem{MacKerellBashfordBellott.1998.All-Atom-Empirical-Potential-for-Molecu%
lar-Modeling}
J.~A.~D. MacKerell, D.~Bashford, M.~Bellott, J.~R.~L.~Dunbrack, J.~D. Evanseck,
  M.~J. Field, S.~Fischer, J.~Gao, H.~Guo, S.~Ha, D.~Joseph-McCarthy,
  L.~Kuchnir, K.~Kuczera, F.~T.~K. Lau, C.~Mattos, S.~Michnick, T.~Ngo, D.~T.
  Nguyen, B.~Prodhom, I.~W.~E.~Reiher, B.~Roux, M.~Schlenkrich, J.~C. Smith,
  R.~Stote, J.~Straub, M.~Watanabe, J.~Wi\'{o}rkiewicz-Kuczera, D.~Yin, and
  M.~Karplus,
\newblock {J. Phys. Chem. B} \textbf{102}, 3586 (1998).

\bibitem{BrooksBrooksMackerell.2009.CHARMM:-The-Biomolecular-Simulation-Progra%
m}
B.~R. Brooks, I.~Brooks, C.~L., J.~Mackerell, A.~D., L.~Nilsson, R.~J.
  Petrella, B.~Roux, Y.~Won, G.~Archontis, C.~Bartels, S.~Boresch, A.~Caflisch,
  L.~Caves, Q.~Cui, A.~R. Dinner, M.~Feig, S.~Fischer, J.~Gao, M.~Hodoscek,
  W.~Im, K.~Kuczera, T.~Lazaridis, J.~Ma, V.~Ovchinnikov, E.~Paci, R.~W.
  Pastor, C.~B. Post, J.~Z. Pu, M.~Schaefer, B.~Tidor, R.~M. Venable, H.~L.
  Woodcock, X.~Wu, W.~Yang, D.~M. York, and M.~Karplus,
\newblock {J. Comp. Chem.}  \textbf{30}, 1545 (2009).

\bibitem{DuanWuChowdhury.2003.A-point-charge-force-field-for-molecular-mechani%
cs}
Y.~Duan, C.~Wu, S.~Chowdhury, M.~C. Lee, G.~M. Xiong, W.~Zhang, R.~Yang,
  P.~Cieplak, R.~Luo, T.~Lee, J.~Caldwell, J.~M. Wang, and P.~Kollman,
\newblock {J. Comp. Chem.} \textbf{24}, 1999 (2003).

\bibitem{Schulz:2009p5308}
R.~Schulz, B.~Lindner, and L.~Petridis,
\newblock {J. Chem. Theory Comput.} \textbf{5}, 2798 (2009).

\bibitem{Feller:1996p5303}
S.~Feller, R.~Pastor, A.~Rojnuckarin, S.~Bogusz, and B.~Brooks,
\newblock {J. Phys. Chem} \textbf{100}, 17011 (1996).

\bibitem{Spohr.1997.Effect-of-Electrostatic-Boundary-Conditions-and-System}
E.~Spohr,
\newblock {J. Chem. Phys.} \textbf{107}, 6342 (1997).

\bibitem{HummerSoumpasisNeumann.1994.Computer-simulation-of-aqueous-Na-Cl-Elec%
trolytes}
G.~Hummer, D.~M. Soumpasis, and M.~Neumann,
\newblock {J. Phys. -- Condens. Matt.} \textbf{6}, A141 (1994).

\bibitem{FennellGezelter.2006.Is-the-Ewald-summation-still-necessary-Pairwise}
C.~J. Fennell and J.~D. Gezelter,
\newblock {J. Chem. Phys.} \textbf{124}, 234104 (2006).

\bibitem{WolfKeblinskiPhillpot.1999.Exact-method-for-the-simulation-of-Coulomb%
ic-systems}
D.~Wolf, P.~Keblinski, S.~R. Phillpot, and J.~Eggebrecht,
\newblock {J. Chem. Phys.} \textbf{110}, 8254 (1999).

\bibitem{WuBrooks.2005.Isotropic-periodic-sum:-A-method-for-the-calculation}
X.~Wu and B.~R. Brooks,
\newblock {J. Chem. Phys.} \textbf{122}, 044107 (2005).

\bibitem{WuBrooks.2008.Using-the-isotropic-periodic-sum-method-to-calculate}
X.~Wu and B.~R. Brooks,
\newblock {J. Chem. Phys.} \textbf{129}, 154115 (2008).

\bibitem{WuBrooks.2009.Isotropic-periodic-sum-of-electrostatic-interactions-fo%
r-polar}
X.~Wu and B.~R. Brooks,
\newblock {J. Chem. Phys.} \textbf{131}, 024107 (2009).

\bibitem{Patra:2003p5327}
M.~Patra, M.~Karttunen, M.~Hyv{\"o}nen, E.~Falck, P.~Lindqvist, and
  I.~Vattulainen,
\newblock {Biophys. J.} \textbf{84}, 3636 (2003).

\bibitem{Patra:2004p5329}
M.~Patra, M.~Karttunen, M.~Hyvonen, E.~Falck, and I.~Vattulainen,
\newblock {J. Phys. Chem. B} \textbf{108}, 4485 (2004).

\bibitem{Patra:2006p5332}
M.~Patra, E.~Salonen, E.~Terama, I.~Vattulainen, R.~Faller, B.~Lee,
  J.~Holopainen, and M.~Karttunen,
\newblock {Biophys. J.} \textbf{90}, 1121 (2006).

\bibitem{ChenKaurWeeks.2004.Connecting-systems-with-short-and-long}
Y.~G. Chen, C.~Kaur, and J.~D. Weeks,
\newblock {J. Phys. Chem. B} \textbf{108}, 19874 (2004).

\bibitem{RodgersWeeks.2008.Local-molecular-field-theory-for-the-treatment}
J.~M. Rodgers and J.~D. Weeks,
\newblock {J. Phys. -- Condens. Matt.} \textbf{20}, 494206 (2008).

\bibitem{HuWeeks.2010.Efficient-Solutions-of-Self-Consistent-Mean-Field}
Z.~Hu and J.~D. Weeks,
\newblock {Phys. Rev. Lett.} \textbf{105}, 140602 (2010).

\bibitem{HansenMcDonald.2006.Theory-of-Simple-Liquids}
J.-P. Hansen and I.~R. McDonald.
\newblock {\emph{Theory of Simple Liquids}}.
\newblock Academic Press, New York, 3$^{rd}$ edition, (2006).

\bibitem{McQuarrie.2000.Statistical-Mechanics}
D.~A. McQuarrie.
\newblock {\emph{Statistical Mechanics}}.
\newblock University Science Books, Sausalito, California, (2000).

\bibitem{RodgersWeeks.2008.Interplay-of-local-hydrogen-bonding-and-long-ranged%
-dipolar}
J.~M. Rodgers and J.~D. Weeks,
\newblock {Proc. Nat. Acad. Sci. USA} \textbf{105}, 19136 -- 19141, (2008).

\bibitem{RodgersKaurChen.2006.Attraction-between-like-charged-walls:-Short-ran%
ged}
J.~M. Rodgers, C.~Kaur, Y.-G. Chen, and J.~D. Weeks,
\newblock {Phys. Rev. Lett.} \textbf{97}, 097801 (2006).

\bibitem{RodgersWeeks.2009.Accurate-thermodynamics-for-short-ranged-truncation%
s-of-Coulomb}
J.~M. Rodgers and J.~D. Weeks,
\newblock {J. Chem. Phys.} \textbf{131}, 244108 (2009).

\bibitem{Nezbeda.2005.Towards-a-Unified-View-of-Fluids}
I.~Nezbeda,
\newblock {Mol. Phys.} \textbf{103}, 59 (2005).

\bibitem{BerendsenGrigeraStraatsma.1987.The-missing-term-in-effective-pair-potentials}
H.~J.~C. Berendsen, J.~R. Grigera, and T.~P. Straatsma,
\newblock {J. Phys. Chem.} \textbf{91}, 6269 (1987).

\bibitem{NikitinLyubartsev.2007.New-six-site-acetonitrile-model-for-simulation%
s-of-liquid}
A.~M. Nikitin and A.~P. Lyubartsev,
\newblock {J.Comp. Chem.} \textbf{28}, 2020 (2007).

\bibitem{SmithYongRodger.2002.DLPOLY:-Application-to-molecular-simulation}
W.~Smith, C.~W. Yong, and P.~M. Rodger,
\newblock {Mol. Sim.} \textbf{28}, 385 (2002).

\bibitem{TakahashiNarumiYasuoka.2010.Cutoff-radius-effect-of-the-isotropic-per%
iodic}
K.~Takahashi, T.~Narumi, and K.~Yasuoka,
\newblock {J. Chem. Phys.} \textbf{133}, 014109 (2010).

\bibitem{ChenWeeks.2006.Local-molecular-field-theory-for-effective}
Y.~G. Chen and J.~D. Weeks,
\newblock {Proc. Nat. Acad. Sci. USA} \textbf{103}, 7560 (2006).

\bibitem{Rodgers.2008.Statistical-Mechanical-Theory-for-and-Simulations-of-Cha%
rged}
J.~M. Rodgers,
\newblock PhD thesis, University of Maryland, http://hdl.handle.net/1903/8561,
  (2008).
  
\bibitem{Mullinax:2009p4696}
J.~W. Mullinax and W.~G. Noid,
\newblock {Phys. Rev. Lett.} \textbf{103}, 198104 (2009).

\bibitem{TaylorLipson.1994.A-Site-Site-Born-Green-Yvon-Equation-for-Hard-Spher%
e}
M.~P. Taylor and J.~E.~G. Lipson,
\newblock {J. Chem. Phys.} \textbf{100}, 518 (1994).


\bibitem{ChandlerPratt.1976.Statistical-Mechanics-of-Chemical-Equilibria-and-I%
ntramolecular}
D.~Chandler and L.~R. Pratt,
\newblock {J. Chem. Phys.} \textbf{65}, 2925 (1976).

\bibitem{Smith.2006.DLPOLY-applications-to-molecular-simulation-II}
W.~Smith,
\newblock {Mol. Sim.} \textbf{32}, 933 (2006).

\end{thebibliography}

\appendix


\section{Yvon-Born-Green (YBG) Equation and Local Molecular Field (LMF) Theory Derived for Small Site-Site Molecules in an External Field \label{app:SimpleYBGandLMF}}
While this paper primarily deals with a uniform site-site molecular
fluid, the derivations of both the YBG hierarchy as well as the LMF
equation are simpler for a general nonuniform system.  We present this derivation
here using a straightforward method that also
introduces the basic site-site notation and ideas
that we will then generalize and apply to the uniform fluid  in
Appendices~\ref{app:YBG} and~\ref{app:LMFbulk},
where careful attention is paid to the distinction between intra- and intermolecular correlations.
We should note that Mullinax and Noid~\cite{Mullinax:2009p4696} have developed a
basis expansion method that can
be used to derive a generalized YBG equation for a variety of molecular systems.

Previous
work~\cite{TaylorLipson.1994.A-Site-Site-Born-Green-Yvon-Equation-for-Hard-Sphere}
for site-site \ybg\ equations begins the derivation by writing the
singlet density for a molecular site in terms of the singlet density
for the entire molecule with a fixed orientation, taking appropriate gradients on either side,
and only then reducing to a site-site representation.  Using the general
formalism developed by Chandler and
Pratt~\cite{ChandlerPratt.1976.Statistical-Mechanics-of-Chemical-Equilibria-and-Intramolecular}
for the partition functions and density distribution functions of
mixtures of site-site molecular models, we may follow a similar path
to the derivation of a general site-site \ybg\ equation.  The formalism
originally was developed to also account for the possibility of
chemical reactions, and since this is not a concern in the inherently
classical systems we study, a few alterations will be made to simplify
notation, with no impact on the meaning of the equations.

The partition function for a mixture of molecular species $M$ with
total sites $n_M$ on each molecule labeled by Greek characters such as
$\xi$ is given below with the position of the $\xi$ site on the
$i^{\rm th}$ molecule of type $M$ given as $\vect{r}_{iM}^{(\xi)}$ and
the positions of all $n_M$ sites on the $i^{\rm th}$ molecule of type $M$
given as $\vect{R_{iM}}$.
\begin{equation}
  Q(\{M\}) = \left( \prod_{M} N_M! \nu_M^{N_M} \prod_{\xi=1}^{n_M} \left(\Lambda_{M}^{(\xi)}\right)^{3N_M}\right)^{-1} \int \, e^{-\beta \hamiltonian} \,\left(\prod_{M,i}d\vect{R_{iM}}\right) 
\end{equation}
where the total potential energy \hamiltonian\ is defined as
\begin{align}
&\hamiltonian = \sum_M \sum_{i=1}^{N_M} \omega_M(\vect{R_{iM}}) + \sum_M \sum_{i=1}^{N_M} \sum_{\xi=1}^{n_M} \phi_{M,\xi}(\vect{r_{iM}^{(\xi)}}) \nonumber \\
& \qquad + \frac{1}{2} \sum_{M}\sum_{M^\prime}\sum_{i=1}^{N_M} \sum_{j=1}^{N_{M^\prime}} (1 -\delta_{MM^\prime}\delta_{ij})\sum_{\xi=1}^{n_M} \sum_{\alpha=1}^{n_{M^\prime}} u_{\xi M \alpha M^\prime}\left(\vdiff{r_{iM}^{(\xi)}}{r_{jM^\prime}^{(\alpha)}}\right).
\end{align}
Here $\nu_M$ is the symmetry number of the molecule.  For example,
for H$_2$O, $\nu=2$ for 2 equivalent orientations, and for CH$_4$,
$\nu=12$ for 12 different equivalent orientations -- 3 equivalent
rotations for each of 4 different C-H bonds fixed in position.  With
symmetry numbers included, each ``equivalent'' atom may be correctly
viewed as a \emph{unique} site.  Thus H$_2$O has 3 sites and CH$_4$
would have 5 sites.  $\Lambda_{M}^{(\xi)}$ is the thermal de
Broglie wavelength for the atom $\xi$ on molecule M.  The factor of
$(1-\delta_{MM^\prime}\delta_{ij})$ ensures that the general pair
interactions $u_{\xi M\alpha M^\prime}$, often taken as a sum of
Coulomb and LJ interactions in CHARMM-like models, arise only for
sites on different molecules.  We will consider modifications
necessary to apply this reasoning to a true CHARMM model for
larger molecules in Appendix~\ref{app:charmm}.

We now write the single-site density distribution function using the
notation $d\overline{\vect{R}}$ to represent all molecular coordinates
$\vect{R}_{iM}$, and ``division'' by $d\vect{r}_{1M}^{(\xi)}$ to indicate
integration over all particle positions except the $\xi$ site on
the 1$^{\rm st}$ molecule of type $M$.  Thus, we have
\begin{equation}
  \rho_{\xi M}^{(1)}(\vect{r}) = \frac{N_M}{Z} \int \, e^{-\beta \hamiltonian} \, \left( \frac{d\overline{\vect{R}}}{d\vect{r_{1M}^{(\xi)}}}\right),
   \label{eq:genericsinglet}
\end{equation}
with $Z_{N}$ the configurational partition function and normalization
constant given by integration over all $\overline{\mathbf R}$. Here,
$\vect{r}$ has replaced $\vect{r_{1M}^{(\xi)}}$ in \hamiltonian.
Now taking the gradient with respect to $\vect{r}$ and using the
equivalence of all molecules of a given type,
\begin{align}
  &  -k_BT\del \rho_{\xi M}^{(1)}(\vect{r}) = \frac{N_M}{Z} \int \,  \left[\del \omega_M(\vect{R_{1M}})\right] e^{-\beta \hamiltonian} \, \left( \frac{d\overline{\vect{R}}}{d\vect{r_{1M}^{(\xi)}}}\right) \nonumber \\
  & \qquad + \left[\del \phi_{M,\xi}(\vect{r})\right] \, \frac{N_M}{Z} \int \,  e^{-\beta \hamiltonian} \, \left( \frac{d\overline{\vect{R}}}{d\vect{r_{1M}^{(\xi)}}}\right)\nonumber \\
  & \qquad + \frac{N_M(N_M-1)}{Z} \int \, \left[ \sum_{\alpha=1}^{n_M} \del u_{\xi M \alpha M}\left(\vdiff{r}{r_{2M}^{(\alpha)}}\right)\right]e^{-\beta \hamiltonian} \, \left( \frac{d\overline{\vect{R}}}{d\vect{r}_{1M}^{(\xi)}}\right) \nonumber \\
  &\qquad + \frac{N_M N_{M^\prime}}{Z} \int \,
  \left[\sum_{M^\prime\neq M} \sum_{\alpha=1}^{n_{M^\prime}} \del
    u_{\xi M \alpha
      M^\prime}\left(\vdiff{r}{r_{1M^\prime}^{(\alpha)}}\right)\right]e^{-\beta
    \hamiltonian} \, \left(
    \frac{d\overline{\vect{R}}}{d\vect{r}_{1M}^{(\xi)}}\right).
    \label{eq:sitesitegen}
\end{align}
We may simplify this site-site molecular \ybg\ 
equation in terms of an intramolecular density distribution function,
$\varrho_M(\vect{R_{M}})$, and a two-point intermolecular site-site
density distribution function, $\rho_{\xi M \alpha
  M^\prime}^{(2)}(\vect{r},\vect{r^\prime})$, specifically defined to
exclude intramolecular site-site correlations.  Here we set
$\vect{R_{1M}}=\vect{R_M}$, $\vect{r_{1M}^{(\xi)}}=\vect{r}$, and
$\vect{r_{2M^\prime}^{(\alpha)}}=\vect{r^\prime}$ in \hamiltonian:
\begin{align}
  \varrho_M(\vect{R}_M) &= \frac{N_M}{Z}\int \,  e^{-\beta \hamiltonian} \, \left( \frac{d\overline{\vect{R}}}{d\vect{R}_{1M}} \right) \\
  \rho_{\xi M \alpha M^\prime}^{(2)}(\vect{r},\vect{r^\prime}) &=
  \frac{N_M (N_{M^\prime} - \delta_{MM^\prime})}{Z} \int \,
  e^{-\beta \hamiltonian} \, \left(
    \frac{d\overline{\vect{R}}}{d\vect{r_{1M}^{(\xi)}}d\vect{r_{2M^\prime}^{(\alpha)}}}
  \right).
\end{align}
Substituting these definitions into equation (\ref{eq:sitesitegen}), we find
\begin{align}
& -k_BT\del\rho_{\xi M}^{(1)}(\vect{r}) = \int \, \left[\del \omega_M(\vect{R_{M}})\right] \varrho_M( \vect{R_M}) \, \left( \frac{d\vect{R_{M}}}{d\vect{r_{M}^{(\xi)}}}\right)  \nonumber \\
& \qquad \qquad +\left[ \del \phi_{M,\xi}(\vect{r}) \right] \rho_{\xi M}^{(1)}(\vect{r}) + \sum_{M^\prime} \sum_{\alpha=1}^{n_{M^\prime}} \int d\vect{r^\prime} \rho_{\xi M \alpha M^\prime}^{(2)}(\vect{r},\vect{r^\prime})\del u_{\xi M \alpha M^\prime}(|\vect{r}-\vect{r^\prime}|).
\end{align}

The sole difference between this equation and the \ybg\ equation for
atomic mixtures is the term involving the gradient of the bonding energy and
the intramolecular density distribution function.  In order to put this exact
\ybg\ equation in a standard form from which the \lmf\ equation is derived,
we divide each side by $\rho_{\xi M}^{(1)}(\vect{r})$, yielding
\begin{align}
  & -k_BT\del \left( \ln \rho_{\xi M}^{(1)}(\vect{r}) \right) = \int \,  \left[\del \omega_M(\vect{R_M})\right] \varrho_{M|\xi}(\vect{R_M}|\vect{r})\,  \left( \frac{d\vect{R_M}}{d\vect{r_{M}^{(\xi)}}}\right) \nonumber \\
  & \qquad \qquad + \del \phi_{M\xi}(\vect{r}) + \sum_{M^\prime}
  \sum_{\alpha=1}^{n_{M^\prime}} \int d\vect{r^\prime} \rho_{\alpha
    M^\prime | \xi M}(\vect{r^\prime}|\vect{r})\del u_{\xi
    M \alpha M^\prime}(|\vect{r}-\vect{r^\prime}|).
    \label{eq:YBGcond}
\end{align}
This division generates conditional densities on the right side
of equation~(\ref{eq:YBGcond}).  Thus
\begin{equation}
\rho_{\alpha M^\prime| \xi M}(\vect{r^\prime}|\vect{r}) \equiv
\rho_{\xi M \alpha M^\prime}^{(2)}(\vect{r},\vect{r^\prime})/
\rho_{\xi M}^{(1)}(\vect{r})
\end{equation}
is an intermolecular conditional density,
proportional to the probability of finding
site $\alpha$ on a molecule of type $M^\prime$ at position $\vect{r^\prime}$ given
that site $\xi$ on a molecule of type $M$ is located at position
$\vect{r}$, and similarly $\varrho_{M|\xi}(\vect{R_M}|\vect{r})$ is the
intramolecular conditional density of a molecular orientation
$\vect{R_M}$ given that site $\xi$ is located at position
$\vect{r}$.

We now derive the LMF equation.  We first consider a general separation of the intermolecular
interactions into short- and long-ranged parts
\begin{equation}
\label{eq:genpotsep}
u_{\alpha M \xi M^\prime}(r)=u_{0,\alpha M \xi M^\prime}(r)+u_{1,\alpha M \xi M^\prime}(r),
\end{equation}
where $u_{1}$ is slowly varying over the range of strong nearest-neighbor interactions.
We seek a mimic system which is composed of molecules with only short-ranged
intermolecular interactions $u_{0,\alpha M \xi M^\prime}(r)$
along with effective single-particle
potentials ${\phi}_{R,\xi M}(\vect{r})$, chosen in principle so that the induced
singlet densities in the full and mimic systems are equal:
\begin{equation}
\label{eq:Nonuniformequaldensities}
\rho^{(1)}_{R,\xi M}(\vect{r};[\phi_R])=\rho^{(1)}_{\xi M}(\vect{r};[\phi]).
\end{equation}
All intramolecular and bonding potentials will be assumed to be the
same in the mimic and full systems.
 
Following the standard path to the  \lmf\ derivation, we take the exact
difference between the \ybg\ equation for the full system and the
\ybg\ equation for a mimic system, assuming the equality of the
singlet density profiles.  After rearrangement we find
\begin{align}
  & \del \phi_{R,\xi M}(\vect{r}) =  \del \phi_{\xi M}(\vect{r}) + \sum_{M^\prime} \sum_{\xi=1}^{n_{M^\prime}} \int d\vect{r^\prime} \rho_{R, \xi M^\prime} (\vect{r^\prime}) \, \del u_{1,\xi M \alpha M^\prime} \left(\vdiff{r}{r^\prime}\right)  \nonumber \\
  & \qquad + \int \left \{ \varrho_{M|\xi}(\vect{R_M}|\vect{r};[\phi]) - \varrho_{R,M|\xi}(\vect{R_M}|\vect{r};[\phi_R]) \right \} \left[\del \omega_M(\vect{R_M})\right] \left( \frac{d\vect{R}_M}{d\vect{r_{M}^{(\xi)}}}\right) \nonumber \\
  & \qquad + \sum_{M^\prime} \sum_{\alpha=1}^{n_{M^\prime}}
  \int d\vect{r^\prime} \left \{ \rho_{\alpha M^\prime | \xi M
    }(\vect{r^\prime}|\vect{r};[\phi]) - \rho_{R, \alpha M^\prime |
      \xi M }(\vect{r^\prime}|\vect{r}; [\phi_R] ) \right \}
\del u_{0,\xi M \alpha
    M^\prime}(|\vect{r}-\vect{r^\prime}|) \nonumber \\
  & \qquad + \sum_{M^\prime} \sum_{\alpha=1}^{n_{M^\prime}} \int
  d\vect{r^\prime} \left \{ \rho_{\alpha M^\prime | \xi
      M}(\vect{r^\prime}|\vect{r}; [\phi])- \rho_{\alpha M^\prime}
    (\vect{r^\prime};[\phi_R]) \right \} \del
  u_{1,\xi M \alpha M^\prime}\left(\vdiff{r}{r^\prime}\right).
  \label{eq:exactdiff}
\end{align}

The above equation is exact but not particularly useful as it stands because of the
appearance of complicated conditional densities on the right hand side.  In order
to yield the LMF equation, we must make three connected and very reasonable
approximations for the integrands of the last three terms based on our chosen forms
for $u_{0}$ and $u_{1}$.
\begin{itemize}
\item {\bf Approximation 1:} The densities of specific molecular orientations will
be well approximated by the mimic system such that 
\begin{equation}
 \varrho_{M|\xi}(\vect{R_M}|\vect{r};[\phi]) \simeq \varrho_{R,M|\xi}(\vect{R_M}|\vect{r};[\phi_R]),
  \label{eq:MolecDensApprox}
\end{equation}
allowing neglect of the integrand involving these functions.  For
small molecules, this seems like an eminently reasonable
approximation, since the prevalence of various relative intramolecular
orientations in both systems will be dominated by the identical short-ranged interactions and the
overall molecular orientation should be quite well approximated given
local short-ranged interactions and the long-ranged orientational
corrections due to \Vr.
\item {\bf Approximation 2:} The product $\left \{ \rho_{\alpha M^\prime |
      \xi M }(\vect{r^\prime}|\vect{r};[\phi]) - \rho_{R, \alpha
      M^\prime | \xi M }(\vect{r^\prime}|\vect{r}; [\phi_R] )
  \right \} \del u_{0,\xi M \alpha
    M^\prime}(|\vect{r}-\vect{r^\prime}|)$ can be neglected.  This
  term probes the difference between the conditional singlet densities
  for the full and mimic systems via convolution with $\del \us$.  The
  integrand will be quickly forced to zero at larger
  $\vdiff{r}{r^\prime}$ by the vanishing gradient of the short-ranged
  \us.  The integrand will also be negligible at small
  $\vdiff{r}{r^\prime}$ since both the full and mimic systems have the
  same strong short-ranged core forces with an appropriately-chosen
  \us, so the density difference inside the curly brackets should then be very small.
\item {\bf Approximation 3:} The final product $\left \{ \rho_{\alpha M^\prime
      | \xi M}(\vect{r^\prime}|\vect{r}; [\phi])- \rho_{\alpha
      M^\prime} (\vect{r^\prime};[\phi_R]) \right \} \del u_{1,\xi
    M \alpha M^\prime}\left(\vdiff{r}{r^\prime}\right)$ can also be
    neglected.  This is due to the fact that difference between the
  conditional singlet density and the singlet density of the
  \emph{full} system will be most substantial for exactly the small
  distances where $u_{1}$ is slowly varying and $\del u_1\left(\vdiff{r}{r^\prime}\right)$ will
  be small. At large separations the conditional singlet density reduces to
  the usual singlet density
  except in special cases like near the critical point, so this term can again be neglected.
  
\end{itemize}
Approximation 1 is the sole new addition as Approximations 2 and 3 are
identical to those required for single site mixtures as detailed in
Ref.~\cite{RodgersWeeks.2008.Local-molecular-field-theory-for-the-treatment}.
However, when these reasonable approximations are employed and 
LMF theory is applied only to the charge-charge interactions of
molecular models so that charge densities can be introduced as
in \cite{RodgersWeeks.2008.Local-molecular-field-theory-for-the-treatment},
we can exactly integrate the remaining terms in
equation~(\ref{eq:exactdiff}) and find the desired  \lmf\ equation for site-site molecular
models:
\begin{align}
  \phi_{R,\xi M}(\vect{r}) &= \phi_{ne,\xi M}(\vect{r}) + q_{\xi M} \Vr \left(\vect{r}\right) \nonumber \\
  \Vr(\vect{r}) &= \mathcal{V}(\vect{r}) + \frac{1}{\epsilon} \int
  d\vect{r}^\prime \, \rho^q_R(\vect{r}^\prime) v_1
  \left(\vdiff{r}{r^\prime}\right).
  \label{eq:LMFsitesite}
\end{align}
Here $\phi_{ne,\xi M}(\vect{r})$ contains all the non-Coulombic parts
of the external field and $\mathcal{V}(\vect{r})$ is the electrostatic potential from
the fixed charge distribution as explained in detail
in \cite{RodgersWeeks.2008.Local-molecular-field-theory-for-the-treatment}.
Each molecular site now moves in a renormalized electrostatic potential
\Vr\ due to an average charge density $\rho_R^q(\vect{r})$ that is
partially contributed to by it and its bound molecular sites.  This
might seem to be a cause for concern, since implementations of Ewald summation
do remove the effect of both the charge itself and these bound
charges~\cite{Smith.2006.DLPOLY-applications-to-molecular-simulation-II}.
However, we argue that this is reasonable since \lmf\ theory convolutes the
\emph{average} charge density, not the instantaneous charge density,
with the slowly-varying long-ranged \vl.

The equation above is \emph{identical} to the mixture \lmf\ equation
as related in previous derivations.  However, the preceding derivation
for small site-site molecules helps us to understand that the use of
the mixture \lmf\ equation for site-site molecules still is grounded
in the \ybg\ equation with solid statistical mechanical
approximations.  It also sets the stage for the notationally more
complex derivations for bulk site-site molecules given below.

\section{Derivation of YBG Equation Appropriate for Uniform Small Site-Site Molecules \label{app:YBG}}
Now we derive the YBG equation for pair distribution functions in a
uniform system of small site-site molecules.  We
first consider a system of only one molecular type in order to focus on the new
features needed to easily separate out contributions from intra- and
intermolecular interactions.  It is straightforward
to generalize these results to a mixture of molecular types as indicated at the end of the
appendix, and this method can also provide an alternate derivation of
equation (\ref{eq:YBGcond}) as well.

Our broad strategy in deriving the YBG equation for site-site pair distribution
functions in uniform fluids uses the equivalent functional forms of the
pair density function and the conditional singlet density.  The
conditional singlet density may be physically interpreted as the
density that would arise if a single site were  fixed at the origin.
In the following derivation, we apply a special external potential in
the Hamiltonian which yields exactly this situation.  Note that this is different
than the standard use of an external potential to represent an entire molecule
with a given orientation fixed at the origin. Due to the
intramolecular correlations, several new terms arise in the YBG
hierarchy.

In a classical system even identical molecules or sites can be treated
as distinguishable.  It will prove useful to generalize the external
fields $\phi^{(\alpha)}$ appearing in the Hamiltonian in
Appendix~\ref{app:SimpleYBGandLMF} by assuming that the system
interacts with a set of external fields
$\overline{\phi}=\{\phi_{i}^{(\alpha)}({\mathbf r}_{i}^{(\alpha)})\}$
that in principle can differ for each site $\alpha$ of each molecule
$i$.  The total potential energy of the nonuniform molecular system
with this very general set of external fields can then be written as:
\begin{equation}
 {\cal U}(\overline{\mathbf R})=\sum_{i=1}^N  \sum_{\alpha=1}^n \phi_{i}^{(\alpha)}
({\mathbf r}_i^{(\alpha)}) +
 \sum_{i=1}^N 
\omega_{M}(\mathbf R_{i}) +\frac{1}{2}\sum_{i=1}^{N} \sum_{j=1}^N (1-\delta_{ij}) \sum_{\alpha=1}^n
\sum_{\xi=1}^n u_{\alpha\xi}(|{\mathbf r}_i^{(\alpha)}-{\mathbf r}^{(\xi)}_{j}|) .
\label{eq:u}
\end{equation}

We first consider molecule-specific distribution functions like 
\begin{equation}
P^{(1)}_{\xi}({\mathbf r}_1^{(\xi)};[\overline{\phi}]) =
\frac{1}{Z_{N}} \int e^{-\beta {\cal U}(\overline{\mathbf R})
} \frac{d\overline{\mathbf R}}{d{\mathbf r}_1^{(\xi)}},
\label{eq:P1}
\end{equation}
the probability density for finding site $\xi$ of particular molecule
$1$ at ${\mathbf r}_1^{(\xi)}$ and will later consider the usual
generic distribution functions like that given in
equation~(\ref{eq:genericsinglet}), which account for the equivalence of
molecules of the same type.  By taking the gradient of equation~(\ref{eq:P1}) we immediately derive the first equation of the specific
YBG hierarchy:
\begin{eqnarray}
-k_B T \nabla_{{\mathbf r}_1^{(\xi)}} P^{(1)}_{\xi}({\mathbf r}_1^{(\xi)};[\overline{\phi}]) &=&
P^{(1)}_{\xi}({\mathbf r}_1^{(\xi)};[\overline{\phi}])
\nabla_{{\mathbf r}_1^{(\xi)}} \phi^{(\xi)}_{1}({\mathbf r}_1^{(\xi)}) \nonumber \\
&&+ \int P_M ({\mathbf R}_{1};[\overline{\phi}])
\nabla_{{\mathbf r}_1^{(\xi)}} \omega_{M}({\mathbf R}_{1})
\,\frac{d{\mathbf R}_{1}}{d{\mathbf r}_1^{(\xi)}} \nonumber \\
&& + \sum_{j=2}^{N} \sum_{\alpha={1}}^n\int P^{(2)}_{\xi\alpha}({\mathbf r}_1^{(\xi)},
{\mathbf r}_j^{(\alpha)};[\overline{\phi}]) \nabla_{{\mathbf r}_1^{(\xi)}}
u_{\xi\alpha}(|{\mathbf r}_1^{(\xi)}-{\mathbf r}_j^{(\alpha)}|) \,d{\mathbf r}_j^{(\alpha)}.
\label{eq:1strhorybg}
\end{eqnarray}

This YBG equation is identical to that derived in
Appendix~\ref{app:SimpleYBGandLMF}, with the important difference that
it does not appeal to the indistinguishability of molecules of the
same type.  This is crucial because the external field we will apply
explicitly fixes one site on a given molecule at the origin.  Here $P_M
({\mathbf R}_{1};[\overline{\phi}])$ in the second term on the right
denotes the $n$-site intramolecular distribution function, defined as
in equation~(\ref{eq:P1}) but with integration over ${\mathbf R}_{1}$
excluded. The integration in the second term is over all ${\mathbf
  R}_{1}$ with site $\xi$ fixed at ${\mathbf r}_1^{(\xi)}$.  Similarly
the definition of $P^{(2)}_{\xi\alpha}({\mathbf r}_1^{(\xi)},{\mathbf
  r}_j^{(\alpha)};[\overline{\phi}])$ excludes integration over
${\mathbf r}_1^{(\xi)}$ and ${\mathbf r}_j^{(\alpha)}$ and involves
sites on different molecules $1$ and $j$.

We want to determine intermolecular site-site pair distribution
functions in the uniform system with $\overline{\phi} =0$:
$P^{(2)}_{\xi\alpha}({\mathbf r}_1^{(\xi)},{\mathbf
  r}_j^{(\alpha)};[\overline{\phi}=0]) = P^{(2)}_{\xi\alpha}(|{\mathbf
  r}_1^{(\xi)}-{\mathbf r}_j^{(\alpha)}|)$. Even with a general
anisotropic $\omega_{M}$, these can depend only on the radial distance
between sites $\xi$ and $\alpha$ on different molecules $1$ and $j$
because of translation invariance and the spherical symmetry of the
intermolecular potential $u_{\alpha\xi}$ in equation~(\ref{eq:u}).

\begin{figure}
  \begin{center}
    \includegraphics[width=3.0in]{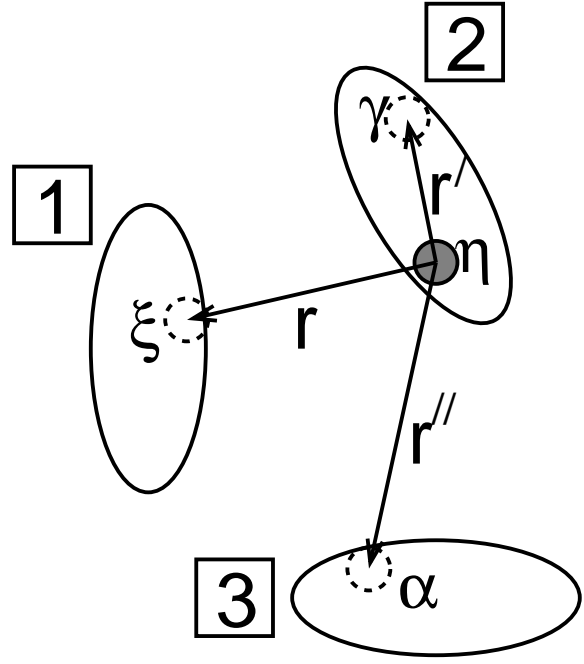}
    \caption{Diagram to show indices and coordinates of various molecular sites, schematically
    represented by small circles, with large ellipses representing a general anisotropic
    molecular bonding potential $\omega_{M}({\mathbf R}_{i})$. Site $\eta$ in molecule 2 is fixed at the origin.}
    \label{fig:indices}
  \end{center}
\end{figure}

We gain information about these uniform system functions by
considering another special case of equation~(\ref{eq:1strhorybg}) where
only a single field $\phi_{2}^{(\eta)}({\mathbf r}_{2}^{(\eta)})$
involving a given site $\eta$ on a particular molecule $2$ is nonzero.
This field  has a special form that confines this site to a very small
spherical region centered about the origin ${\mathbf 0}$. Thus
$\phi_{2}^{(\eta)}({\mathbf r}_{2}^{(\eta)})=\infty$ for $|{\mathbf
  r}_{2}^{(\eta)}| >\epsilon$ and is zero otherwise and we are
interested in the limit $\epsilon \rightarrow 0^{+}$. All other
$\phi_{j}^{(\alpha)}$ are zero.  In order to aid in visualization of
the various sites and molecules, the basic inter-relation of site
indices used in this appendix are shown in Fig.~\ref{fig:indices}.

Note that the nonzero field $\phi_{2}^{(\eta)}({\mathbf r}_{2}^{(\eta)})$
only appears implicitly in
equation~(\ref{eq:1strhorybg}) through its effect on the distribution functions and that this field
fixes only the single site $\eta$ of molecule $2$
at the origin, and not the orientation of the entire molecule.
In the limit $\epsilon \rightarrow 0^{+}$, $P^{(1)}_{\xi}({\mathbf r}_1^{(\xi)};[\phi_{2}^{(\eta)}])$ in
equation~(\ref{eq:1strhorybg}) reduces to a conditional singlet density with site
$\eta$ of molecule $2$ fixed at the origin. Taking account spherical symmetry we write this as
\begin{equation}
P^{(1)}_{\xi}({\mathbf r}_1^{(\xi)};[\phi_{2}^{(\eta)}])=P^{(1)} _{\xi|\eta}(r|{\mathbf 0}),
\label{eq:Pcon}
\end{equation}
where we set ${\mathbf r}={\mathbf r}_1^{(\xi)}$ and note
that $P^{(1)} _{\xi|\eta}$ can depend only on the magnitude
$r$ of  ${\mathbf r}$. The bar before the subscript $\eta$ and the
argument ${\mathbf 0}$ on the right side indicates a conditional density with site
$\eta$ in constrained molecule 2 fixed at the origin.  By translational invariance the specific pair distribution function
$P^{(2)}_{\xi\eta}(r)$ in the uniform system equals $V^{-1}$
times the corresponding specific conditional singlet density in equation~(\ref{eq:Pcon}).
and we will use this equality later to determine uniform system pair distribution functions.

The nonuniform pair distribution functions in equation~(\ref{eq:1strhorybg}) can be similarly
rewritten in this special case. In particular, the pair distribution function involving
another site $\gamma \neq \eta$ on constrained molecule $2$ can be written as
\begin{equation}
\label{eq:P2con}
P^{(2)}_{\xi\gamma}({\mathbf r}_1^{(\xi)},{\mathbf r}_2^{(\gamma)};[\phi_{2}^{(\eta)}])=
P^{(2)}_{\xi\gamma|\eta}({\mathbf r},{\mathbf r}^{\prime}|{\mathbf 0})
\end{equation}
where we set ${\mathbf r}_2^{(\gamma)}={\mathbf r}^{\prime}$.
In this and the following appendix we will
generally use a single prime to denote coordinates on the constrained
molecule.  $P^{(2)}_{\xi\gamma|\eta}$ is strongly affected by the
fixed site and the short-ranged intramolecular interaction
$\omega_{M}$ in equation~(\ref{eq:u}) and vanishes for large $|{\mathbf
  r}^{\prime}|$. This is even more true for the distribution function
$P^{(2)}_{\xi\eta}({\mathbf r}_1^{(\xi)},{\mathbf
  r}_2^{(\eta)};[\phi_{2}^{(\eta)}])$, which has the limiting form as
$\epsilon \rightarrow 0^{+}$
\begin{equation}
P^{(2)}_{\xi\eta}({\mathbf r}_1^{(\xi)},{\mathbf r}_2^{(\eta)};[\phi_{2}^{(\eta)}])=
P^{(1)}_{\xi|\eta}(r|{\mathbf 0})\delta({\mathbf r}_2^{(\eta)}-{\mathbf 0}).
\label{eq:P1delta}
\end{equation}
Both these distribution functions are very different from those involving any site $\alpha$ on an unconstrained third molecule, which takes the form
\begin{equation}
P^{(2)}_{\xi\alpha}({\mathbf r}_1^{(\xi)},{\mathbf r}_3^{(\alpha)};[\phi_{2}^{(\eta)}])=
P^{(2)}_{\xi\alpha|\eta}({\mathbf r},{\mathbf r}^{\prime\prime}|{\mathbf 0})
\end{equation}
where we set ${\mathbf r}_3^{(\alpha)}={\mathbf r}^{\prime\prime}$, and generally use double primes to denote
coordinates of unconstrained molecules.
In contrast to equation~(\ref{eq:P2con}), this does not vanish for large $|{\mathbf r}^{\prime\prime}|$, where
it reduces to a product of conditional single particle functions for large $|\vect{r}-\vect{r}^{\prime\prime}|$.  See Fig.~\ref{fig:indices}.

We also define an induced single particle interaction on site $\xi$ associated with the pair
potential from fixed site $\eta$ at ${\mathbf 0}$:
\begin{equation}
{\phi}_{\xi|\eta}(r)\equiv u_{\xi\eta}(r)
\label{eq:phitilde}
\end{equation}
and rewrite equation~(\ref{eq:1strhorybg}) using the new notation in this special case. Separating terms
involving constrained molecule 2 from those that involve other unconstrained molecules, we get
\begin{eqnarray}
-k_BT\nabla_{\mathbf r} P^{(1)}_{\xi|\eta}(r|{\mathbf 0}) &=&
P^{(1)}_{\xi|\eta}(r|{\mathbf 0})
\nabla_{\mathbf r} {\phi}_{\xi|\eta}(r) \nonumber \\
&&+ \int P_{M|\eta} ({\mathbf R}|{\mathbf 0})
\nabla_{\mathbf r} \omega_{M}({\mathbf R})
\,\frac{d{\mathbf R}}{d{\mathbf r}} \nonumber \\
&& + \sum_{\gamma \neq {\eta}}^n\int
P^{(2)}_{\xi\gamma|\eta}({\mathbf r},{\mathbf r^{\prime}}|{\mathbf 0})
\nabla_{{\mathbf r}}
u_{\xi\gamma}(|{\mathbf r}-{\mathbf r}^{\prime}|) \,d{\mathbf r}^{\prime} \nonumber \\
&& + (N-2)\sum_{\alpha={1}}^n\int
P^{(2)}_{\xi\alpha|\eta}({\mathbf r},{\mathbf r}^{\prime\prime}|{\mathbf 0})
\nabla_{{\mathbf r}}
u_{\xi\alpha}(|{\mathbf r}-{\mathbf r}^{\prime\prime}|) \,d{\mathbf r}^{\prime\prime}.
\label{eq:condybg1}
\end{eqnarray}
Using Eqs.\ (\ref{eq:P1delta}) and (\ref{eq:phitilde}), the first term
on the right side of equation~(\ref{eq:condybg1}) is generated by the
$\alpha=\eta$ and $j=2$ term in equation~(\ref{eq:1strhorybg}), where the
pair potential from the fixed site $\eta$ acts like an effective
external field on site $\xi$.  We have used the equivalence of all
molecules except $1$ and $2$ in the last term in equation~(\ref{eq:condybg1}).

To get to the final form useful for LMF theory we divide by $P^{(1)}_{\xi|\eta}(r|{\mathbf 0})$ and
introduce the usual generic distribution functions. Thus the distribution function for finding
site $\xi$ of any other molecule at ${\mathbf r}$ is
\begin{equation}
\rho^{(1)}_{\xi|\eta}(r|{\mathbf 0})\equiv(N-1)P^{(1)}_{\xi|\eta}(r|{\mathbf 0})
\end{equation}
Similarly the generic distribution function involving three distinct molecules in the last line of equation~(\ref{eq:condybg1}) is
\begin{equation}
\rho^{(2)}_{\xi\alpha|\eta}({\mathbf r},{\mathbf r}^{\prime\prime}|{\mathbf 0})\equiv
(N-1)(N-2)P^{(2)}_{\xi\alpha|\eta}({\mathbf r},{\mathbf r}^{\prime\prime}|{\mathbf 0})
\end{equation}
Division by $\rho^{(1)}_{\xi|\eta}(r|{\mathbf 0})$ will yield a density conditioned by $\xi$ as well, defined by
\begin{equation}
\rho^{(1)}_{\alpha|\eta\xi}({\mathbf r}^{\prime\prime}|{\mathbf 0},{\mathbf r})\equiv
\rho^{(2)}_{\xi\alpha|\eta}({\mathbf r},{\mathbf r}^{\prime\prime}|{\mathbf 0})/
\rho^{(1)}_{\xi|\eta}(r|{\mathbf 0})
\label{eq:condendef}
\end{equation}

The remaining distribution functions in equation~(\ref{eq:condybg1})
involve sites on only two molecules and have very different forms
strongly influenced by the intramolecular interaction $\omega_{M}$. We
again use the symbol $\varrho$ to emphasize this point and define
generic functions
\begin{equation}
\varrho_{M|\eta} ({\mathbf R}|{\mathbf 0})\equiv
(N-1)P_{M|\eta} ({\mathbf R}|{\mathbf 0})
\end{equation}
and
\begin{equation}
\varrho^{(2)}_{\xi\gamma|\eta}({\mathbf r},{\mathbf r}^{\prime}|{\mathbf 0})\equiv
(N-1)P^{(2)}_{\xi\gamma|\eta}({\mathbf r},{\mathbf r}^{\prime}|{\mathbf 0}).
\end{equation} 
Densities conditioned on $\xi$ as well are similarly defined as in equation~(\ref{eq:condendef}).

Using this notation in equation~(\ref{eq:condybg1}) we arrive at the desired final form
for the first equation of the site-site molecular YBG hierarchy, with site $\eta$ of a
particular molecule fixed at the origin:
\begin{eqnarray}
-k_B T\nabla_{\mathbf r} \ln\rho^{(1)}_{\xi|\eta}(r|{\mathbf 0}) &=&
\nabla_{\mathbf r} {\phi}_{\xi|\eta}(r) \nonumber \\
&&+ \int \varrho_{M|\eta\xi} ({\mathbf R}|{\mathbf 0},{\mathbf r})
\nabla_{\mathbf r} \omega_{M}({\mathbf R})
\,\frac{d{\mathbf R}}{d{\mathbf r}} \nonumber \\
&& + \sum_{\gamma \neq {\eta}}^n\int
\varrho^{(1)}_{\gamma|\eta\xi}({\mathbf r}^{\prime}|{\mathbf 0},{\mathbf r})
\nabla_{{\mathbf r}}
u_{\xi\gamma}(|{\mathbf r}-{\mathbf r}^{\prime}|) \,d{\mathbf r}^{\prime} \nonumber \\
&& + \sum_{\alpha={1}}^n\int
\rho^{(1)}_{\alpha|\eta\xi}({\mathbf r}^{\prime\prime}|{\mathbf 0},{\mathbf r})
\nabla_{{\mathbf r}}
u_{\xi\alpha}(|{\mathbf r}-{\mathbf r}^{\prime\prime}|) \,d{\mathbf r}^{\prime\prime}.
\label{eq:condybg2}
\end{eqnarray}
Note that this YBG equation is nearly identical to the equation
derived in Appendix~\ref{app:SimpleYBGandLMF}.  An important
additional contribution arises due the correlations between the site
$\xi$ and the various sites $\gamma$ present on the \emph{same
  molecule} as the site $\eta$ fixed at the origin.

It is straightforward to extend this approach to a general mixture of molecular species.
With obvious generalizations of notation we find for site $\xi$ of species $M$ with
site $\eta$ of a different molecule of a possibly
different species $M^{\prime}$ fixed at ${\mathbf 0}$:
\begin{eqnarray}
-k_B T\nabla_{\mathbf r} \ln\rho^{(1)}_{\xi M|\eta M^{\prime}}(r|{\mathbf 0}) &=&
\nabla_{\mathbf r} {\phi}_{\xi M|\eta M^{\prime}}(r) \nonumber \\
&&+ \int \varrho_{M|\eta M^{\prime}\xi M} ({\mathbf R}|{\mathbf 0},{\mathbf r})
\nabla_{\mathbf r} \omega_{M}({\mathbf R})
\,\frac{d{\mathbf R}}{d{\mathbf r}} \nonumber \\
&& + \sum_{\gamma \neq {\eta}}^{n_{M^{\prime}}}\int
\varrho^{(1)}_{\gamma M^{\prime}|\eta M^{\prime}\xi M}({\mathbf r}^{\prime}|{\mathbf 0},{\mathbf r})
\nabla_{{\mathbf r}}
u_{\xi M\gamma M^{\prime}}(|{\mathbf r}-{\mathbf r}^{\prime}|) \,d{\mathbf r}^{\prime} \nonumber \\
&& + \sum_{M^{\prime\prime}}\sum_{\alpha={1}}^{n_{M^{\prime\prime}}}\int
\rho^{(1)}_{\alpha M^{\prime\prime}|\eta M^{\prime}\xi M}
({\mathbf r}^{\prime\prime}|{\mathbf 0},{\mathbf r})\nabla_{{\mathbf r}}
u_{\xi M\alpha M^{\prime\prime}}(|{\mathbf r}-{\mathbf r}^{\prime\prime}|) \,d{\mathbf r}^{\prime\prime}.
\label{eq:condybg2mix}
\end{eqnarray}

\section{Derivation of LMF Equation Appropriate for Uniform Small Site-Site Molecules \label{app:LMFbulk}}
We now derive the LMF equations appropriate for a uniform mixture of
site-site molecules, using the exact YBG
equation~(\ref{eq:condybg2mix}).  The
basic strategy follows that for the molecular system considered in
Appendix~\ref{app:SimpleYBGandLMF}.  We again consider a general
separation of the intermolecular interactions into short- and
long-ranged parts, as in equation~(\ref{eq:genpotsep}) such that the mimic
system will have only short-ranged intermolecular interactions along
with effective single-particle interactions ${\phi}_{R,\xi M|\eta
  M^\prime}(r)$ associated with the fixed site at the origin.  These
effective interactions are again chosen in principle so that the
induced densities in the full and mimic systems are equal:
\begin{equation}
\label{eq:equaldensities}
\rho^{(1)}_{R,\xi M|\eta M^{\prime}}(r|{\mathbf 0})=\rho^{(1)}_{\xi M|\eta M^{\prime}}(r|{\mathbf 0}).
\end{equation}
All intramolecular and bonding potentials will be assumed to be the
same in the mimic and full systems.  In Appendix~\ref{app:charmm}, we
generalize to instances of larger molecules where long-ranged
interactions might exist between sites on the same molecule.

Following the standard path to \lmf\ derivation, we take the exact
difference between the \ybg\ equation for the full system and the
\ybg\ equation for a mimic system in a restructured field for which
equation~(\ref{eq:equaldensities}) holds.  Since we already must include
subscripts for the fixed site and two other sites, for simplicity of
notation we will first consider a single component site-site molecular
system.

Thus, using equations (\ref{eq:equaldensities}) and (\ref{eq:condybg2}) we have exactly
\begin{eqnarray}
\del _{\mathbf r} [{\phi}_{R,\xi |\eta }(r)- {\phi}_{\xi |\eta }(r)]&=&
 \int \bigg \{ \varrho_{M|\eta\xi} ({\mathbf R}|{\mathbf 0},{\mathbf r})-
\varrho_{R,M|\eta\xi} ({\mathbf R}|{\mathbf 0},{\mathbf r})\bigg  \}
\nabla_{\mathbf r} \omega_{M}({\mathbf R})
\,\frac{d{\mathbf R}}{d{\mathbf r}} \nonumber \\
&& + \sum_{\gamma \neq {\eta}}^n\int
\bigg \{\varrho^{(1)}_{\gamma|\eta\xi}({\mathbf r}^{\prime}|{\mathbf 0},{\mathbf r})
-\varrho^{(1)}_{R,\gamma|\eta\xi}({\mathbf r}^{\prime}|{\mathbf 0},{\mathbf r})\bigg \}
\nabla_{{\mathbf r}}
u_{0,\xi\gamma}(|{\mathbf r}-{\mathbf r}^{\prime}|) \,d{\mathbf r}^{\prime} \nonumber \\
&& + \sum_{\alpha={1}}^n\int
\bigg \{\rho^{(1)}_{\alpha|\eta\xi}({\mathbf r}^{\prime\prime}|{\mathbf 0},{\mathbf r})
-\rho^{(1)}_{R,\alpha|\eta\xi}({\mathbf r}^{\prime\prime}|{\mathbf 0},{\mathbf r})\bigg \}
\nabla_{{\mathbf r}}
u_{0,\xi\alpha}(|{\mathbf r}-{\mathbf r}^{\prime\prime}|) \,d{\mathbf r}^{\prime\prime}  \nonumber \\
&& + \sum_{\gamma \neq {\eta}}^n\int
\bigg \{\varrho^{(1)}_{\gamma|\eta\xi}({\mathbf r}^{\prime}|{\mathbf 0},{\mathbf r})
-\varrho^{(1)}_{\gamma|\eta}({r}^{\prime}|{\mathbf 0})\bigg \}
\nabla_{{\mathbf r}}
u_{1,\xi\gamma}(|{\mathbf r}-{\mathbf r}^{\prime}|) \,d{\mathbf r}^{\prime}  \nonumber \\
&& + \sum_{\alpha={1}}^n\int
\bigg \{\rho^{(1)}_{\alpha|\eta\xi}({\mathbf r}^{\prime\prime}|{\mathbf 0},{\mathbf r})
-\rho^{(1)}_{\alpha|\eta}({r}^{\prime\prime}|{\mathbf 0})\bigg \}
\nabla_{{\mathbf r}}
u_{1,\xi\alpha}(|{\mathbf r}-{\mathbf r}^{\prime\prime}|) \,d{\mathbf r}^{\prime\prime} \nonumber \\
&& + \sum_{\gamma \neq {\eta}}^n\int
\varrho^{(1)}_{R,\gamma|\eta}({r}^{\prime}|{\mathbf 0})
\nabla_{{\mathbf r}}
u_{1,\xi\gamma}(|{\mathbf r}-{\mathbf r}^{\prime}|) \,d{\mathbf r}^{\prime}  \nonumber \\
&& + \sum_{\alpha={1}}^n\int
\rho^{(1)}_{R,\alpha|\eta}({r}^{\prime\prime}|{\mathbf 0})
\nabla_{{\mathbf r}}
u_{1,\xi\alpha}(|{\mathbf r}-{\mathbf r}^{\prime\prime}|) \,d{\mathbf r}^{\prime\prime}
\label{eq:exactsub}
\end{eqnarray}
 
As a consequence of our judicious choice of \us\ and \ul, all the
integrals involving terms with large curly brackets vanish to a good approximation.
The first, third, and fifth terms with curly brackets all may be
neglected by Approximations 1-3 as detailed in
Appendix~\ref{app:SimpleYBGandLMF}.  We now must employ two related
approximations leading to cancellation of the intramolecular
correlations functions. 
\begin{itemize}
\item {\bf Approximation 4} The product $\left
    \{\varrho^{(1)}_{\gamma|\eta\xi}({\mathbf r}^{\prime}|{\mathbf
      0},{\mathbf r}) -\varrho^{(1)}_{R,\gamma|\eta\xi}({\mathbf
      r}^{\prime}|{\mathbf 0},{\mathbf r})\right \} \nabla_{{\mathbf
      r}} u_{0,\xi\gamma}(|{\mathbf r}-{\mathbf r}^{\prime}|)$ will be
  approximately zero.  The logic here is virtually identical to that
of Approximation 2.  Given 
rigid or even flexible bonds between intramolecular sites $\gamma$ and
$\eta$ we expect the matchup between densities in the short-ranged system and the full system to be even better at short distances, leading to an even stronger cancellation.
\item {\bf Approximation 5} The product $\left
    \{\varrho^{(1)}_{\gamma|\eta\xi}({\mathbf r}^{\prime}|{\mathbf
      0},{\mathbf r})
    -\varrho^{(1)}_{\gamma|\eta}({r}^{\prime}|{\mathbf 0})\right \}
  \nabla_{{\mathbf r}} u_{1,\xi\gamma}(|{\mathbf r}-{\mathbf
    r}^{\prime}|)$ will also be approximately zero, for reasons
  similar to Approximation 3.  In fact, the intramolecular conditional
  density profiles should be less sensitive to the presence of a site
  on another molecule for many configurations.  At small separations,
  the cancellation due to the slowly-varying nature of \ul\ will still hold.
\end{itemize}

Thus we see that while more intramolecular terms must cancel in the
derivation of the LMF equation, exactly the same line of logic is
followed as in Appendix~\ref{app:SimpleYBGandLMF}.
Using (\ref{eq:equaldensities}) and setting the first five integral
terms to zero as justified in the previous discussion, we arrive at
the site-site LMF equations for each combination of fixed site $\eta$
and mobile site $\xi$:
\begin{eqnarray}
\label{eq:gensitesite}
{\phi}_{R,\xi |\eta }(r)- {\phi}_{\xi |\eta }(r)&=&
\sum_{\gamma \neq {\eta}}^n\int
\varrho^{(1)}_{R,\gamma|\eta}({r}^{\prime}|{\mathbf 0})
u_{1,\xi\gamma}(|{\mathbf r}-{\mathbf r}^{\prime}|) \,d{\mathbf r}^{\prime} \nonumber \\
&& + \sum_{\alpha={1}}^n\int
\rho^{(1)}_{R,\alpha|\eta}({r}^{\prime\prime}|{\mathbf 0})
u_{1,\xi\alpha}(|{\mathbf r}-{\mathbf r}^{\prime\prime}|) \,d{\mathbf r}^{\prime\prime} + C.
\end{eqnarray}
In the above equation, there are terms due to intramolecular
sites as well as sites on other molecules.  The portion due to
intramolecular sites does not imply an action of \phiR\ on
intramolecular sites but rather includes the \emph{effect} of these
intramolecular sites on sites of other molecules. This set of equations for each
choice of $\xi$ and $\eta$ has the simplest form
possible with a general separation of the pair interactions $u_{\xi\eta}$ into
short- and long-ranged parts.

However, as discussed in detail
in~\cite{RodgersWeeks.2008.Local-molecular-field-theory-for-the-treatment},
LMF theory takes a particularly simple and powerful form when it is
applied only to Coulomb interactions and all charges are separated
using the same $\sigma$, as we do in this paper.  Charge densities rather than individual
molecular site densities can then be naturally introduced, as shown below.
Furthermore, these new
equations based on charge densities are not only simpler, but likely lead to an
even stronger overall cancellation of terms than argued for each of
the individual terms previously. 

We may write the long-ranged Coulomb part of the specific
intermolecular pair interactions as
\begin{equation}
\label{eq:coulombu1}
u_{1,\alpha M \xi M^\prime}(r) = \frac{q_{\alpha M} q_{\xi M^\prime}}{\epsilon} \vl
\end{equation}
and as before the short-ranged core interactions will be defined as
$u_{0, \alpha M \xi M^\prime}(r) = u_{\alpha M \xi M^\prime}(r)
- u_{1,\alpha M \xi M^\prime}(r)$ and will encompass all \lj-like
interactions as well as the usual Coulomb core  \vs\ terms.
In particular, using equation~(\ref{eq:phitilde}), the induced interaction from the
fixed site can be written as
\begin{equation}
\label{eq:induced}
{\phi}_{\xi |\eta }(r)={\phi}_{ne,\xi |\eta }(r) +
\frac{q_{\xi} q_{\eta}}{\epsilon}[\vs + \vl ],
\end{equation}
where ${\phi}_{ne,\xi |\eta }(r)$ contains all non-electrostatic (usually LJ) pair interactions
between sites $\xi$ and $\eta$.

The relevant spherically symmetric charge distribution arising from
the constrained molecule with
site $\eta$ fixed at the origin is given by
\begin{equation}
\label{eq:moleculechargedensity}
\varrho^{q}_{R,M|\eta}({r}|{\mathbf 0}) \equiv
q_{\eta}\delta ({\mathbf r}-{\mathbf 0}) +
\sum_{\gamma \neq {\eta}}^n
q_{\gamma} \varrho^{(1)}_{R,\gamma|\eta}({r}|{\mathbf 0}).
\end{equation}
For rigid molecules like SPC/E water,
$\varrho^{q}_{R,M|\eta}({r}|{\mathbf 0})$ can be determined in advance
and expressed solely in terms of sums of $\delta$-functions as discussed in
equations (\ref{eq:Hsite}) and (\ref{eq:Osite}).

In general, the total induced equilibrium charge density for a site $\eta$
fixed at the origin is then
\begin{equation}
\label{eq:totalchargedensity}
\rho_{R,\tot}^{q}({r |{\mathbf 0}}) \equiv
\varrho^{q}_{R,M|\eta}({r}|{\mathbf 0}) +
\sum_{\alpha={1}}^n
q_{\alpha}\rho^{(1)}_{R,\alpha|\eta}({r}|{\mathbf 0}),
\end{equation}
where the second term is the contribution to the charge density
from the other unconstrained mobile molecules.

Using (\ref{eq:coulombu1}), we see equation~(\ref{eq:gensitesite})
can now be written in the compact form
\begin{equation}
\label{eq:inducedR}
{\phi}_{R,\xi |\eta }(r)={\phi}_{ne,\xi |\eta }(r) +
q_{\xi}  {\mathcal{V}_{R|\eta}}(r),
\end{equation}
where ${\mathcal{V}_{R|\eta}}(r)$ is the restructured electrostatic potential
induced by the fixed site $\eta$ and the other associated intramolecular sites.
This satisfies the site-site Coulomb LMF equation
\begin{equation}
{\mathcal{V}_{R|\eta}}(r) = \frac{q_\eta}{\epsilon}\vs + \frac{1}{\epsilon} \int
  d\vect{r}^\prime \, \rho_{R,\tot}^{q}({r^{\prime} |{\mathbf 0}}) v_1
  \left(\vdiff{r}{r^\prime}\right).
  \label{eq:LMFgeneralcoulomb}
\end{equation}

We may also define a restructured potential $\mathcal{V}_{R1|\eta}$
containing only the long-ranged components of the potentials as
\begin{equation} {\mathcal{V}_{R1|\eta}}(r) \equiv
  \mathcal{V}_{R|\eta}(r) - \frac{q_\eta}{\epsilon}\vs =
  \frac{1}{\epsilon} \int d\vect{r}^\prime \,
  \rho_{R,\tot}^{q}({r^{\prime} |{\mathbf 0}}) v_1
  \left(\vdiff{r}{r^\prime}\right).
  \label{eq:LMFgeneralVRlongcoulomb}
\end{equation}
This is the restructured Gaussian-smoothed electrostatic
potential induced by the fixed charge from the site
$\eta$ at the origin, where $\rho^q_{R,\tot}$ is the total equilibrium charge
density due to the fixed charge, the intramolecular charge density, as
well as the fully mobile charges on other molecules.

\section{Treatment for Non-Uniform and Uniform Larger CHARMM-like
  Molecules \label{app:charmm}}

While the use of Approximation 1 in previous derivations of LMF theory
might suggest that our findings are invalid for larger molecules
defined by CHARMM- or AMBER-like potentials, this is not the case.
Far less restrictive approximations than the equivalence of the
whole molecule density functions may be derived by the usual postulates of a
bonding potential form where only sites separated by 1, 2, or 3
consecutive bonds in a molecule may experience a special local bonding interaction. 
Pairs of intramolecular sites with larger separations interact
only through spherically symmetric
(usually Coulomb and LJ) pair potentials.

Rather than proceeding through logic identical to that found in the
previous appendices, we instead outline the approximations necessary.
Then we briefly describe features of the LMF equations valid for such larger molecules in a
general external field and in a uniform fluid. We seek to emphasize that LMF theory
is equally valid for large molecular models typically employed, based
on physically reasonable approximations.

A separate appendix dealing with \ybg\ equations and \lmf\ equations
for larger site-site molecules is necessary because in most simulation
potentials, such as those defined by the {\sc
  charmm}~\citep{MacKerellBashfordBellott.1998.All-Atom-Empirical-Potential-for-Molecular-Modeling}
and {\sc
  amber}~\citep{DuanWuChowdhury.2003.A-point-charge-force-field-for-molecular-mechanics}
parameter sets, the potential energy due to ``intermolecular''
interactions (\lj\ interactions and point charge interactions) is not
written as distinct summations over molecules and their intramolecular
sites.  Rather the \lj\ and charge interaction contribution to
\hamiltonian\ is a sum over all sites separated by at least three
bonds (\ie\ excluding atoms bonded or connected via angle bending).

The expression for the partition function $Q$ does not change, but
\hamiltonian\ does.  Specifically, we decompose the general $\omega_M$ into a set
of bonds $\omega^{(b)}_{\alpha \gamma M}$, bond angles
$\omega^{(a)}_{\alpha \gamma \delta M}$, and bond dihedrals
$\omega^{(d)}_{\alpha \gamma \delta \zeta M}$ connecting
appropriate sets of neighboring sites.  We also introduce a
bonding matrix $B_M(\xi,\alpha)$ for each species $M$ which is 1 if sites
$\xi$ and $\alpha$ can be connected by two consecutive bonds and 0
otherwise.  With this notation we write
\begin{align}
&\hamiltonian = \sum_M \sum_{i=1}^{N_M} \sum_{\xi=1}^{n_M} \phi_{\xi M}(\vect{r_{iM}^{(\xi)}}) + \sum_M \sum_{i=1}^{N_M} \sum_{\alpha-\gamma} \omega^{(b)}_{\alpha \gamma M}\left(\vdiff{r_{iM}^{(\alpha)}}{r_{iM}^{(\gamma)}}\right) \nonumber \\
& \qquad + \sum_M \sum_{i=1}^{N_M} \sum_{\alpha-\gamma-\delta} \omega^{(a)}_{\alpha \gamma \delta M}\left(\vect{r}_{iM}^{(\alpha)},\vect{r}_{iM}^{(\gamma)}, \vect{r}_{iM}^{(\delta)}\right) \nonumber \\
& \qquad + \sum_M \sum_{i=1}^{N_M} \sum_{\alpha-\gamma-\delta-\zeta} \omega^{(d)}_{\alpha\gamma \delta \zeta M}\left(\vect{r}_{iM}^{(\alpha)},\vect{r}_{iM}^{(\gamma)}, \vect{r}_{iM}^{(\delta)}, \vect{r}_{iM}^{(\zeta)} \right) \nonumber \\
& \qquad + \frac{1}{2} \sum_{M}\sum_{M^\prime}\sum_{i=1}^{N_M} \sum_{j=1}^{N_{M^\prime}} \sum_{\xi=1}^{n_M} \sum_{\alpha=1}^{n_{M^\prime}} \left (1 -\delta_{MM^\prime}\delta_{ij} B_M(\xi,\alpha) \right) u_{\xi M \alpha M^\prime}\left(\vdiff{r_{iM}^{(\xi)}}{r_{jM^\prime}^{(\alpha)}}\right).
\end{align}
\hamiltonian\ written in this way is virtually identical to the small
site-site molecular \hamiltonian\ other than the decomposition of the
bonding potentials and the allowance for intermolecular-like
interactions between sufficiently separated sites within a single
molecule.  The first three $\omega$ terms are for bond vibrations, angle
vibrations, and dihedral rotations of two bonds around a connecting
bond.  Technically, these usually depend on only $r$, $\theta$, and
$\phi$ respectively, but we include positions for generality and for
ease in taking gradients in deriving the appropriate YBG and LMF
equations.  These sums are understood to count sets of atoms connected
via bond, angular, or torsional potentials only once.  One
complication for the {\sc amber} force field is that non-bonded
interactions are scaled down for 1-4 (dihedral) pairs.  \lj\
interactions for 1-4 pairs are divided by 2.0 and Coulomb interactions
are divided by 1.2.  We will not address this complication, but it
conceivably could be included in the $B_M(\alpha,\gamma)$ formalism by
introducing matrix elements accounting for these scalings.  The new all-atom
force field for {\sc charmm} does not scale the Coulomb interactions
for 1-4 pairs.

Based on the form of the potential, it is quite logical that
Approximation~1 presented in Appendix~\ref{app:SimpleYBGandLMF} now
becomes a series of approximations related to particles connected via
bonding potentials.  Following the same path for LMF derivation in
Appendix~\ref{app:SimpleYBGandLMF}, we find that the \emph{weaker}
conditions for accuracy replacing Approximation 1 are:
\begin{align}
  & \bullet \text{for sites $\alpha$ and $\gamma$ connected via bonds,} \nonumber \\
  & \qquad \qquad    \varrho^{(2)}\left(\vect{r^{(\alpha)}},\vect{r^{(\gamma)}};[\phi]\right) \simeq \varrho_R^{(2)}\left(\vect{r^{(\alpha)}},\vect{r^{(\gamma)}};[\phiR]\right) \label{eq:BondReq} \\
  & \bullet \text{for three sites $\alpha$, $\gamma$, and $\delta$ connected via a bond angle,} \nonumber \\
  & \qquad \qquad    \varrho^{(3)}\left(\vect{r^{(\alpha)}},\vect{r^{(\gamma)}},\vect{r^{(\delta)}};[\phi]\right) \simeq \varrho_R^{(3)}\left(\vect{r^{(\alpha)}},\vect{r^{(\gamma)}}\vect{r^{(\delta)}};[\phiR]\right) \label{eq:AngleReq} \\
  & \bullet \text{and for sites $\alpha$, $\gamma$, $\delta$, and $\zeta$ involved in dihedral rotations,} \nonumber \\
  & \qquad \qquad
  \varrho^{(4)}\left(\vect{r^{(\alpha)}},\vect{r^{(\gamma)}},\vect{r^{(\delta)}},\vect{r^{(\zeta)}};[\phi]\right)
  \simeq
  \varrho_R^{(4)}\left(\vect{r^{(\alpha)}},\vect{r^{(\gamma)}}\vect{r^{(\delta)}},\vect{r^{(\zeta)}};[\phiR]\right). \label{eq:TorsionReq}
\end{align}
These approximations are much more easily supported by mimic systems
with reasonably small \sig.  This \sig\ may have to be on the order of
1-4 distances since 1-4 pairs have Coulomb interactions.  In general
though, we expect that \lmf\ theory can be applied in standard
biomolecular all-atomistic simulations with reasonable success
with a $\sigma$ spanning only a few bond lengths rather than an entire
biomolecular radius, as suggested by our results for acetonitrile in
the main text.

Provided that these approximations hold, we find exactly the same
\lmf\ equation for a nonuniform system.  Analysis for the bulk uniform
fluid becomes more challenging as we must in principle consider all three-particle
combinations of the three sites -- $\alpha$, $\xi$, and the fixed site
$\eta$ -- where $\alpha$ and $\xi$ interact via their pair potential.
This involves a wide range of permutations across different molecules,
resulting in a larger number of terms that have to cancel in the
derivation of the LMF equation, but again the final equation is
essentially identical with the same underlying physical intuition.  In
fact, for these large molecules the attractiveness of treating solely
charge-charge interactions via LMF theory becomes apparent.  Tracking
the net charge density profile about a site is far more manageable than
tracking all possible site-site density profiles.

\end{document}